\documentclass[aps,nofootinbib,prc,twocolumn,superscriptaddress,showpacs,floatfix]{revtex4-1}
\usepackage{graphicx,amssymb}
\usepackage[fleqn]{amsmath}
\usepackage{amsfonts}
\usepackage{dcolumn}% Align table columns on decimal point                      
\usepackage{bm}% bold math                                                      
\usepackage{braket}
\usepackage{color}
\usepackage{multirow}
\usepackage{MnSymbol}

\definecolor{dgreen}{rgb}{0.0,0.5,0.0}

\newcommand{\sfrac}[2]{{\textstyle\frac{#1}{#2}}}
\newcommand{\half}{\textstyle\frac{1}{2}}

\newcommand{\DE}{\Delta^{(3)}_E}
\newcommand{\Dr}{\Delta^{(3)}_r}
\newcommand{\De}{\Delta^{\rm ee}_E}
\newcommand{\Drr}{\delta\langle r^2\rangle}

\renewcommand{\vec}[1]{\mbox{\boldmath $#1$}}

\begin{document}

\title{Towards the global description of nuclear charge radii: exploring the Fayans energy density functional}

\author{P.-G. Reinhard}
\affiliation{Institut f\"ur Theoretische Physik II, Universit\"at Erlangen-N\"urnberg,
D-91058 Erlangen, Germany}

\author{W. Nazarewicz}
\affiliation{Department of Physics and Astronomy and FRIB Laboratory,
Michigan State University, East Lansing, Michigan  48824, USA}
\affiliation{Institute of Theoretical Physics, Faculty of Physics,
University of Warsaw, Warsaw, Poland}

\date{\today}

\begin{abstract}
	\begin{description}
		\item[Background]
			Binding energies and charge radii are fundamental properties of atomic nuclei. When inspecting their particle-number dependence, both quantities exhibit pronounced odd-even staggering. While the odd-even effect in binding energy can be  attributed to nucleonic pairing, the origin of staggering in charge radii  is less straightforward to ascertain.
		\item[Purpose]
			In this work, we study the odd-even effect in binding energies and charge radii, and systematic behavior of differential radii, to identify the  underlying components of the effective nuclear interaction.
		\item[Method]
			We apply nuclear density functional theory using a family of 
			Fayans and Skyrme energy density functionals
            fitted to similar datasets but using different  optimization protocols.  We inspect various correlations between 
			differential charge radii, odd-even staggering in energies and radii, and nuclear matter properties. Detailed analysis is carried out for medium-mass and heavy semi-magic nuclei with a particular focus on the Ca chain.
		\item[Results]
			By making the surface and pairing terms  dependent on density  gradients, the Fayans functional offers the superb simultaneous description of odd-even staggering effects in energies and charge radii. Conversely, when the data on differential radii are added to the pool of fit-observables, the coupling constants determining the strengths of the gradient terms of Fayans functional are increased by orders of magnitude.  
			The Skyrme functional optimized in this work with the generalized Fayans pairing term offers results of similar quality. We quantify these findings by performing correlation analysis based on the statistical linear regression technique. The nuclear matter parameters characterizing Fayans and Skyrme functionals optimized to similar datasets are fairly close.
		\item[Conclusion]
            Fayans paring functional, with its generalized density dependence, significantly improves description of charge radii in odd and even nuclei. Adding differential charge radii to the set of fit-observables  in the optimization protocol is helpful for both description of radii and for improving pairing functional. In particular,  Fayans functional constrained in this way is capable of explaining charge radii in the even-even Ca isotopes. However, in order to obtain good description  of differential radii data in both  medium-mass and heavy nuclei, an $A$-dependent scaling of  Fayans pairing functional is still needed. Various extensions of the current model are envisioned that carry out a promise for the global description.
	\end{description}
\end{abstract}

\maketitle

\section{Introduction}

Charge radii of atomic nuclei are key observables that can probe
properties of of nuclear force and nuclear many-body dynamics
\cite{Hagen16}. They also carry fundamental information about the
saturation density of symmetric nuclear matter \cite{ReiNaz16}. The
local fluctuations (i.e., rapid changes as a function of particle number) in measured charge radii signal structural evolution
effects, such as shell and subshell closures \cite{ang13}, shape
deformations \cite{Sun17}, and configuration mixing \cite{DeWitte}.
Another key structural indicator is the odd-even staggering of charge
radii along isotopic chains.  In particular, the intricate behavior of
charge radii along  the Ca chain -- the almost equal values of charge
radii in $^{40}$Ca and $^{48}$Ca, an appreciable odd-even staggering,
and unexpectedly large charge radius in $^{52}$Ca \cite{GarciaRuiz16}
-- constitute a long-standing challenge for nuclear theory
\cite{GarciaRuiz16,Rossi15,Minamisono16}.

The goal of this study is to understand differential charge radii
within nuclear density functional theory (DFT) \cite{(Ben03)}, which
is a tool of choice for  global microscopic studies of nuclei throughout
the chart of nuclides.  In nuclear DFT, effective inter-nucleon
interaction is represented by the energy density functional (EDF)
adjusted to experimental data and often also to selected nuclear
matter parameters.  While the commonly used energy density functionals
offer a very reasonable description of charge radii
\cite{Patyk99,Goriely09,Kortelainen_2010,Utama16} they often miss
local fluctuations and dramatically underestimate odd-even effect
in radii. In this context, the EDF developed by S.A. Fayans and
collaborators \cite{Smirnov88,Fay94,Borzov1996,Fayans1998,Fayans2000}
stands out as it has demonstrated a rare ability to describe charge
radii in isotopic chains of semi-magic spherical nuclei, including the
challenging Ca chain. A notable achievement of the Fayans model has
been to explain the odd-even staggering effect in charge radii in
terms of the density-dependent nucleonic pairing
\cite{Fay1994,FayZ96,Fayans2000}.

In this work, we carry out detailed analysis of the Fayans energy density functional, paying attention to its unique features, in particular its density dependence and the role played by surface and pairing 
gradient terms. We vary the optimization strategies to achieve unbiased comparison of odd-even staggering of binding  energies and charge radii. By means of the statistical covariance technique, we quantify the intricate relation between the pairing functional and  charge radii.

This article is organized as follows: Section \ref{theory} contains the description of the models and methods used. In particular, it defines the Fayans functional,  datasets employed in various optimization variants, and fitting methodologies employed in this work. The results  are contained in Sec.~\ref{results}. Finally, the summary and outlook are given in Sec.~\ref{conclusions}.

\section{Theoretical framework}\label{theory}

In nuclear DFT, the total binding energy of the nucleus is given by
\begin{equation}
E =  \int 
    \mathcal{E}(\vec{r}) d^{3}\vec{r} 
\end{equation}
where $\mathcal{E}$ is the local EDF  that is supposed to be
a real, scalar, time-even, and isoscalar function of local densities 
and currents.

\subsection{Local densities}

Canonical Hartree-Fock-Bogoliubov (HFB) wave functions and occupation
amplitudes uniquely define the one-body density matrix. In this work,
we only consider time-reversal-invariant states and thus need
only time-even densities. These are \cite{(Ben03),Engel_1975,Per04}:
particle density $\rho_t$, kinetic density $\tau_t$, and spin-orbit
current $\vec{J}_t$, where the isospin index $t$ labels isoscalar ($t
= 0$) and isovector ($t = 1$) densities. For instance, the isoscalar
and isovector particle densities are:
\begin{equation}
\rho_0=\rho_n+\rho_p,~~\rho_1=\rho_n-\rho_p.
\end{equation}
If pairing correlations are present, local pairing densities $\breve\rho_p$ and $\breve\rho_n$  appear as well. In this study, for the purpose of the optimization and statistical analysis, we replace the  full HFB problem with the Hartree-Fock (HF)+BCS problem.

\subsection{The Skyrme functional}

The Skyrme EDF can be decomposed into the kinetic term, Skyrme
interaction term (isoscalar and isovector), pairing EDF, Coulomb term,
and additional corrections, such as the center-of-mass (c.m.) term
\cite{(Ben03),Kluepfel_2009,Kortelainen_2010}. The Skyrme interaction
EDF is ${\cal E}_{\mathrm{Sk}}={\cal E}_{\mathrm{Sk},0}+{\cal
  E}_{\mathrm{Sk},1}$, where
\begin{eqnarray}
{\cal E}_{\mathrm{Sk},t} &=& C_t^{\rho\rho}(\rho_0) \rho_t^2 
  + C_t^{\rho\tau} \rho_t\tau_t   + C_t^{\rho\Delta\rho} \rho_t\Delta\rho_t
\nonumber \\ && 
  + C_t^{\rho \nabla J}  \rho_t\bm{\nabla}\cdot\bm{J}_t +  C_t^{J^2} \bm{J}_t^2,
\label{Skyrme}
\end{eqnarray}
with the coupling constants 
$C_t^{\rho\rho}$ containing an additional dependence on the isoscalar 
density:
\begin{equation}
  C_t^{\rho\rho}(\rho_0)
  =  
  C_{t0}^{\rho\rho}
  + C_{t{\rm D}}^{\rho\rho} \rho_0^\alpha. 
\label{ddep}
\end{equation}
The tensor spin-orbit terms $\propto\bm{J}_t^2$ are neglected
here, i.e., we assume $C_t^{J^2}=0$.

In this study, the Coulomb Hartree term is 
calculated exactly using the proton density $\rho_p$:
\begin{equation}
 {E}_{\rm C}  =  e^2 \int d^3r \, d^3r' \rho_p(\vec{r})
              \frac{1}{|\vec{r}-\vec{r}'|} \rho_p(\vec{r}').
\label{Ecoul}
\end{equation}
The exchange term is computed within the standard Slater 
approximation: 
\begin{equation}
{\cal E}_{\rm C,ex}
=  -\sfrac{3}{4} e^2\left(\frac{3}{\pi}\right)^{1/3}\rho_p^{4/3}.
\label{Slater}  
\end{equation}
The c.m. correction 
$E_\mathrm{cm}=-\langle\hat{P}_\mathrm{cm}^2\rangle/(2mA)$
is added to the total energy after the mean-field equations have been
solved.
The pairing EDF is described by the density-dependent pairing term
${\cal E}_{\mathrm{pair}}={\cal E}_{\mathrm{pair},p}+ {\cal E}_{\mathrm{pair},n}$
\cite{Dob01,Dob02a}: 
\begin{equation}
{\cal E}_{\mathrm{pair},q}= {1\over 4} V_{{\rm pair},q}
\left(1 - \frac{\rho_0}{\rho_{\rm pair}}\right) {\breve\rho_q}^2~~~~~(q=p,n).
\label{vpair}
\end{equation}
We note that the pairing EDF in the Skyrme model has in general
different coupling constants for protons and neutrons, see discussion
in Ref.~\cite{Bertsch09}. In the limit of $\rho_{\rm pair}\rightarrow
\infty$, Eq.~(\ref{vpair}) represents volume pairing  and
  $\rho_{\rm pair}=0.16\,\mathrm{fm}^{-3}$ corresponds to what is
  called surface pairing.  The functional SV-min \cite{Kluepfel_2009}
has a mixed pairing with
$\rho_\mathrm{pair}=0.21159$\,fm$^{-3}$.

\subsection{The Fayans functional}
\label{sec:Fayfunc}

Compared to the Skyrme functional, Fayans FaNDF$^0$ \cite{Fayans1998,Tolok15}, DF3 \cite{Fay94,Kromer95,Horen96,Fayans2000},     and DF3-a \cite{Tolokonnikov2010}  EDFs have a more complex dependence on particle densities that stems from a fractional
form of their density-dependent couplings, novel  folding/density-gradient terms, and
Coulomb-nuclear correlation term. The kinetic energy and Coulomb Hartree terms
of Fayans EDF are exactly the same as in the Skyrme model.
The Fayans interaction functional is usually written in terms of dimensionless densities
\begin{equation}
x_t=\frac{\rho_t}{\rho_{\rm sat}},~~~x_{\rm pair}=\frac{\rho_0}{\rho_{\rm pair}},
\end{equation}
where $\rho_{\rm sat}$ and ${\rho_{\rm pair}}$ are scaling parameters
of Fayans EDF. In the Fayans model, $\rho_{\rm sat}$ is interpreted as
the saturation density of symmetric nuclear matter with Fermi energy
$\varepsilon_F=(9\pi/8)^{2/3}\hbar^2/2mr_s^2$ and the Wigner-Seitz
radius $r_s= (3/4\pi\rho_{\rm sat})^{1/3}$. Note that we have to
distinguish between the  parameter $\rho_{\rm sat}$, which is an
  input to the model and the equilibrium density $\rho_{\rm eq}$,
which is result of optimization and characterizes Fayans
EDF. While these two quantities are close, they are not identical.

In this work, we study Fayans functional in the form of
FaNDF$^0$ as its surface energy is directly expressed through local
densities.  However,  we  re-optimize its parameters under various
  conditions. Thus we distinguish between the ``FaNDF$^0$
  functional'' and the ``FaNDF$^0$
  parametrization,'' where the latter is the FaNDF$^0$ functional with the original
   model parameters of Ref.~\cite{Fayans1998}. The FaNDF$^0$ EDF can be decomposed into volume,
surface, and spin-orbit terms,
\begin{equation}
{\cal E}_{{\rm Fy}}
   =
   {\cal E}_{{\rm Fy}}^\mathrm{v}(\rho)
   +{\cal E}_{{\rm Fy}}^\mathrm{s}(\rho)
   +{\cal E}_{{\rm Fy}}^\mathrm{ls}(\rho,\vec{J}).
\label{EFay}
\end{equation}
The volume term $ {\cal E}_{{\rm Fy}}^\mathrm{v}$ is defined as
Pad\'e approximant:
\begin{equation}
 {\cal E}_{{\rm Fy}}^\mathrm{v}=
 \sfrac{1}{3}\varepsilon_F\rho_{\rm sat}
\left[
{a_+^\mathrm{v}}
  \frac{1\!-\!{h_{1+}^\mathrm{v}}x_0^{{\sigma}}}
       {1\!+\!{h_{2+}^\mathrm{v}}x_0^{{\sigma}}}x_0^2
  +
  {a_-^\mathrm{v}}
  \frac{1\!-\!{h_{1-}^\mathrm{v}}x_0}
       {1\!+\!{h_{2-}^\mathrm{v}}x_0}x_1^2.
\right]
\label{EFay-dens}
\end{equation}
Such density dependence in the volume term had also been studied
  in the context of Skyrme EDF's \cite{Erler10} and found to make not
  much difference as compared to the form (\ref{ddep}). The important
  new aspect is that the surface term $ {\cal E}_{{\rm
    Fy}}^\mathrm{s}$ has also the form of a Pad\'e approximant
involving the gradient of density:
\begin{equation}
{\cal E}_\mathrm{Fy}^\mathrm{s}
  = 
  \sfrac{1}{3}\varepsilon_F\rho_{\rm sat}
  \frac{a_+^\mathrm{s}r_s^2(\vec{\nabla} x_0)^2}
       {1+h_{+}^\mathrm{s}x_0^\sigma
        +h_{\nabla}^\mathrm{s}r_s^2(\vec{\nabla} x_0)^2}.
\label{EFay-grad} 
\end{equation}
Similar as in the Skyrme case \cite{Ben07a,Kor14a}, the spin-orbit term  ${\cal E}_{{\rm Fy}}^\mathrm{ls}$ of Fayans functional is derived  from zero-range
two-body spin-orbit and tensor interactions \cite{Kim92,Kromer95,Horen96,Tolokonnikov2010,Gnezdilov14}. For time-even spherical nuclei, it can be written as:
\begin{equation}
{\cal E}_\mathrm{Fy}^\mathrm{ls}
  = \frac{4\varepsilon_F r_s^2}{3\rho_{\rm sat}}
  \left(\kappa\rho_0\vec{\nabla}\cdot\vec{J}_0
  +
 \kappa'\rho_1\vec{\nabla}\cdot\vec{J}_1  
 +g \vec{J}_0^2 + g'\vec{J}_1^2
 \right).
\label{eq:EFy-ls}
\end{equation}
Again, we ignore here the tensor contributions ($g=g'=0$) as in
the original FaNDF$^0$. The remaining term is identical to
that of the Skyrme functional (\ref{Skyrme}) if one identifies
$C_0^{\rho\nabla J}=\frac{4\varepsilon_F r_s^2}{3\rho_{\rm
    sat}}\kappa$ and $C_1^{\rho\nabla J}=\frac{4\varepsilon_F
  r_s^2}{3\rho_{\rm sat}}\kappa'$.

The Coulomb exchange energy of Fayans functional contains an additional  Coulomb-nuclear correlation term:
\begin{equation}
{\cal E}_{\rm C,ex} =
  -\sfrac{3}{4}e^2\left(\frac{3}{\pi}\right)^{1/3}\rho_p^{4/3}
  (1-{h_\mathrm{C}}x_0^{{\sigma}}).
\label{Ecoulex}
\end{equation}
Finally, the pairing functional of the Fayans model goes beyond the density-dependent ansatz (\ref{vpair}):
\begin{equation}
  {\cal E}_{\mathrm{Fy},q}^\mathrm{pair}=
  \sfrac{2{\varepsilon_F}}{3\rho_{\rm sat}}
 {\breve\rho_q}^2
  \left[f_\mathrm{ex}^\xi
       +h_+^\xi x_{\rm pair}^{\gamma}
       +h_\nabla^\xi r_s^2 (\vec{\nabla} x_{\rm pair})^2\right].
\label{eq:ep2}
\end{equation}
This  pairing functional  is supposed to effectively account for  the coupling to surface vibrations; it has a surface character and  contains the novel  density-gradient term, which is essential for explaining the odd-even staggering in $r_{\rm ch}$ \cite{FayZ96,Fayans2000}.

Following the original FaNDF$^0$ definitions \cite{Fayans1998}, we use
$\hbar^2/2m_p=20.749811$\,MeV\,fm$^2$,
$\hbar^2/2m_n=20.721249$\,MeV\,fm$^2$, $e^2=1.43996448$\,MeV\,fm,
$\rho_{\rm sat}=0.16$\,fm$^{-3}$, and $\sigma=1/3$.  As in
Ref.~\cite{Fayans1998}, we also take $\rho_{\rm pair}=\rho_{\rm
  sat}$. In the FaNDF$^0$ parametrization, the surface-energy
parameter $h_{+}^\mathrm{s}$ was assumed to be equal to
$h_{2+}^\mathrm{v}$. Initially, we released this condition and kept it
as a free parameter. It turned out, however, that the
$h_{+}^\mathrm{s}$ is poorly constrained  by our datasets,
i.e., the associated uncertainties are large. We found that the value
$h_{+}^\mathrm{s}=0$ yields more robust fits than the standard
FaNDF$^0$ value, and we adopted it in our work.  For the same reason,
since the Coulomb-nuclear correlation term hardly impacts the
optimization results, we put $h_\mathrm{C}=0$. Finally, we took the
exponent $\gamma=2/3$ in (\ref{eq:ep2}) as in DF3-a as it has been
shown advantageous for reproducing differential radii
\cite{Fayans2000,Minamisono16,Saperstein2016}. The c.m. correction has
been ignored in the original Fayans model. We include it in this
work for the sake of comparison with Skyrme results (except for
  the original FaNDF$^0$ parametrization); this correction is of
minor importance for the quantities discussed here.
 
\subsection{Mixed Skyrme/Fayans functional}

Skyrme EDF and FaNDF$^0$ EDF differ in two respects. First, the
mean-field part has a different density dependence, particularly in
the surface term (\ref{EFay-grad}), and second, the FaNDF$^0$ pairing
functional has a gradient term, which is absent in the Skyrme model.
Changing two features at once can blur comparisons. Thus we also consider
as intermediate step  a mixed functional, which takes the mean
field part from Skyrme EDF and pairing part from FaNDF$^0$ EDF.  We refer to 
such a  mixed functional as ``Sk+PFy''. The mixed
functional helps in two ways. Comparing Skyrme EDF with Sk+PFy
explores the Fayans pairing functional, while comparing Sk+PFy with
full FaNDF$^0$ EDF highlights the role of the gradient term in Fayans surface functional.  

\subsection{Observables studied}

The basic observables calculated in DFT  are
binding energy $E_{\rm B}$ and local particle densities $\rho_t$.  The
charge density $\rho_\mathrm{ch}$ can be directly obtained from
$\rho_p$ and $\rho_n$ by correcting for proton and neutron form
factors, and the spin-orbit contribution. The key parameters of the
charge density are r.m.s. charge radius $r_\mathrm{ch}$, diffraction
(or box-equivalent) radius $R_\mathrm{diff}$, and surface thickness
$\sigma_{\rm ch}$ \cite{Fri82a}.  In the following, we shall study
isotopic trends of binding energy and charge radii. In particular, we
are going to investigate differential mean-square (ms) charge radii,
which, for a given isotope, are defined as:
\begin{equation}\label{rdiff}
  \delta \langle r^2\rangle^{A,A'}
  =  \langle r_{\rm ch}^2\rangle^{A'} - \langle r_{\rm ch}^2\rangle^{A}. 
\end{equation}
To assess odd-even staggering of charge radii and binding energies, we study
three-point differences (either isotopic or isotonic):
\begin{eqnarray}
  \Dr(A)
&  = &
  \half\left(r_{{\rm ch},A+1}-2r_{{\rm ch},A} + r_{{\rm ch},A-1}\right),\label{D3r}
\\
   \DE(A)
&  = &
  \half\left(E_{{\rm B},A+1}-2E_{{\rm B},A} + E_{{\rm B},A-1}\right).\label{D3E}
\end{eqnarray}
We shall also consider  three-point binding energy differences involving ground states of even-even nuclei with the same $Z$ or $N$:
\begin{equation}\label{Eee}
  \De(A)
  =
    \half\left({E_{{\rm B},A+2} -2E_{{\rm B},A}+ E_{{\rm B},A-2}}\right).
\end{equation}
For open-shell systems, $\De$ is proportional to the inverse of the pairing rotational moment of inertia, i.e., it is an excellent indicator of nucleonic pairing \cite{Hinohara16}.
For the calibration of the spin-orbit
functional, we also look at differences of single-particle
energies, $\varepsilon_\mathrm{ls}$. 

Nuclear matter properties (NMP) in symmetric homogeneous matter
characterize the properties of a given functional. Here, in addition to the equilibrium density, $\rho_{\rm eq}$, and  energy-per-nucleon of symmetric nuclear matter at the equilibrium, $E/A$,  we will investigate the following NMP: incompressibility $K$ and effective
mass $m^*/m$ characterizing the isoscalar response; and  symmetry energy $J$,
slope of symmetry energy  $L$, and  Thomas-Reiche-Kuhn sum-rule enhancement
 $\kappa_\mathrm{TRK}$ characterizing the isovector response, see
Refs.~\cite{Kluepfel_2009,Kortelainen_2010,Nazarewicz2014} for definitions.
  The  sum-rule enhancement $\kappa_{\rm TRK}$ is
an alternative  way to parametrize the isovector effective mass.
 Those NMP can be conveniently used to  characterize results obtained under different  optimization strategies.

\subsection{Optimization strategies and variants}\label{variants}

The free parameters of Fayans EDF need to be constrained by experiment. These parameters should be global  in the
sense that they should provide a reasonable description of finite nuclei and extended nucleonic matter \cite{Fayans1998}. The situation
resembles the Skyrme EDF  optimization strategy \cite{(Ben03)},
where the least-squares method \cite{Bev69aB,Bra97aB} has become the
most widely used approach, see, e.g.,
\cite{Fri86a,Sam02a,Kortelainen_2010,Dob14a}. Here, we are going to adopt this strategy for
tuning  Fayans EDF to the global set of data and to compare it 
 with a Skyrme functional tuned in the same way. To that end, we define a global quality measure
\begin{equation}
  \chi^2(\mathbf{p})
  =
  \sum_{n\in\mathrm{Obs.}}
  \frac{(\mathcal{O}^\mathrm{th}_n(\mathbf{p})
         -\mathcal{O}^\mathrm{exp}_n)^2}
       {\Delta^2\mathcal{O}_n},
\label{eq:chi2}
\end{equation}
where the sum runs over all fit-observables $\mathcal{O}_n$, $\mathcal{O}^\mathrm{th}_n$ are predicted values,
$\mathcal{O}^\mathrm{exp}_n$ are  experimental values, and
$\Delta\mathcal{O}_n$ are adopted errors chosen to regulate the
relative weights of the different observables. The $\chi^2$
is a function of the model parameters $\mathbf{p}$ through the
parameter dependence of $\mathcal{O}^\mathrm{th}_n(\mathbf{p})$.
The optimal parameter set $\mathbf{p}_0$ is the one which minimizes
$\chi^2$, i.e.
$\mathcal{O}^\mathrm{th}_n(\mathbf{p}_0)\leq\mathcal{O}^\mathrm{th}_n(\mathbf{p})$
for all $\mathbf{p}$.

%%%%%% 
\begin{table}[!htb]
\caption{The datasets used to constrain Skyrme and Fayans EDFs
  optimized in this work. The basic dataset of \cite{Kluepfel_2009} is
  considered in all cases; hence, it is not mentioned explicitly. For
  instance, the dataset $\Delta E^{\rm ee}$ includes the basic  dataset
  in addition to the specific data listed below.
The numbers in brackets (energies in MeV and radii in fm) indicate the adopted errors in $\chi^2$
(\ref{eq:chi2}). The adopted errors for $\Delta E^{\rm oe}$ were
chosen in accordance with the same choice for the spectral
gaps in Ref.~\cite{Kluepfel_2009}. The adopted errors for 
$\Delta E^{\rm ee}$ were derived by  estimating the effect
of ground state correlations as in  \cite{Kluepfel_2008}. Experimental nuclear masses and charge radii were taken from Refs.~\cite{NUBASE17} and \cite{ang13}, respectively.}
\begin{ruledtabular}
\begin{tabular}{ll}
 Dataset  & \hspace{2.2cm}{Fit-observables} \\
  \hline \\[-8pt]
{\bf basic:} & dataset of SV-min \cite{Kluepfel_2009}: 
 $E_{\rm B}$, $R_\mathrm{diff}$
 $r_\mathrm{ch}$, $\sigma_{\rm ch}$, $\varepsilon_\mathrm{ls}$ 
\\[4pt]
$\bm{\Delta E^{\rm oe}:}$ &  
neutron $\DE$ (\ref{D3E}) in:
$^{44}$Ca\,(0.24), $^{44}$Ca\,(0.36), \\\
& $^{122}$Sn\,(0.36), $^{124}$Sn\,(0.36),
$^{126}$Sn\,(0.24), $^{128}$Sn\,(0.24), \\
& $^{204}$Pb\,(0.24),
$^{206}$Pb\, (0.36),  $^{210}$Pb\,(0.36);  
\\
&
proton $\DE$ (\ref{D3E}) in:
 $^{86}$Er\,(0.36), $^{88}$Sr\,(0.24), \\
 & $^{90}$Zr\,(0.12),  $^{92}$Mo\,(0.24),  $^{94}$Ru\,(0.24),  $^{136}$\,Xe(0.24), \\
 & $^{138}$Ba\,(0.24), $^{140}$Ce\,(0.24), $^{142}$Nd\,(0.24),  \\
 & $^{144}$Sm\,(0.24),  $^{146}$Gd\,(0.24),  $^{148}$Dy\,(0.24);
\\[4pt]
$\bm{\Delta E^{\rm ee}:}$ 
& neutron $\De$ (\ref{Eee}) in:
$^{44}$Ca\,(0.12), $^{118}$Sn\,(0.36), \\
& $^{120}$Sn\,(0.36), $^{122}$\,Sn(0.13), $^{124}$Sn(0.24); 
\\
& proton $\De$ (\ref{Eee}) in:
 $^{36}$Kr\,(0.36), $^{88}$Sr\,(0.36), \\
 & $^{90}$Zr\,(0.24), $^{92}$Mo\,(0.12), $^{94}$\,Ru(0.24), $^{136}$Xe\,(0.24), \\
 & $^{138}$Ba\,(0.24),  $^{140}$Ce\,(0.24),  $^{142}$Nd\,(0.24),\\
 & $^{214}$Ra\,(0.24), $^{216}$Hg\,(0.24)
\\[4pt]
$\bm{\Delta r:}$ & $\delta \langle r^2\rangle$ (\ref{rdiff}) for Ca
isotopes:
$\delta \langle r^2\rangle^{48,40}$\,(0.008),\\
& $\delta \langle r^2\rangle^{48,44}$\,(0.008), 
$\delta \langle r^2\rangle^{52,48}$\,(0.02);
\\[4pt]
$\bm{\Delta r^{\rm oe}:}$ & odd-even staggering:
 $\delta \langle r^2\rangle^{43,44}$(Ca)\,(0.002),\\
&  $\delta \langle r^2\rangle^{119,120}$(Sn)\,(0.002).
\end{tabular}
\end{ruledtabular}
\label{fitobservables}
\end{table} 
%%%%
To optimize Skyrme and Fayans functionals, we employed several
datasets containing various combinations of fit-observables. They are
listed in Table~\ref{fitobservables}, and the parametrizations
resulting from different combinations of datasets are defined in
Table~\ref{funcionalnames}.
%%%%%% 
\begin{table}[!htb]
\caption{\label{funcionalnames} Energy density functionals optimized in this work and the associated fit-observables as defined in  Table~\ref{fitobservables}. 
The basic SV-min dataset of \cite{Kluepfel_2009} is
  considered in all cases.
 All Fy functionals employ the Fayans functional of  Sec.~\ref{sec:Fayfunc} except for 
Fy(${h_\nabla^s}$=0) and  Fy(${h_\nabla^\xi}$=0), in which the surface and pairing gradient
terms are put to zero, respectively.
Sk+PFy mixes the Skyrme functional in the particle-hole channel
 with the Fayans pairing functional (\ref{eq:ep2}) with $\gamma=1$.
In all cases, the  c.m. correction is added.}
  \begin{ruledtabular}
\begin{tabular}{l|ccc}
EDF  & $\Delta E^{\rm oe}$  & $\Delta r$ & $\Delta r^{\rm oe}$ 
\\
 \hline \\[-8pt]
{\bf Fy(std)}
  & +  & $-$   & $-$ 
\\
{\bf Fy(nogap)} & $-$  & $-$ & $-$ 
\\
{\bf Fy($\bm{\Delta r}$)}
 & +   & + & $-$  
\\
{\bf Fy($\bm{\Delta r,\Delta r^{\rm oe}}$)}
 & +   & + & + 
\\
{\bf Fy($\bm{h_\nabla^\xi}$=0)} &  +   & + & $-$  
\\
{\bf Fy($\bm{h_\nabla^s}$=0)} &  +   & + & $-$  
\\
{\bf Sk+PFy}  & +  &  +  & $-$ 
\\
{\bf Sk+PFy(std)}  & +  &  $-$  & $-$ 
\end{tabular}
\end{ruledtabular}
\end{table} 
%%%

Having optimized the functionals, we use the resulting covariance matrices to carry out 
the  correlation analysis as explained,
e.g., in Refs.~\cite{Naz10a,Dob14a,Erl14b}. In particular, we employ a  dimensionless
product-moment correlation coefficient~\cite{Bra97aB}:
\begin{equation}
  {r}_{AB}
  =
  \frac{|\overline{\Delta A\,\Delta B}|}
       {\sqrt{\overline{\Delta A^2}\;\overline{\Delta B^2}}},
\label{correlator}
\end{equation}
which measures the covariance between two observables ${A}$ and
${B}$. A value ${r}_{AB}=1$ means fully correlated, 
  ${r}_{AB}=-1$ fully anti-correlated, and ${r}_{AB}=0$ uncorrelated.
In linear least squares regression, the square of the correlation
coefficient $r_{AB}^2$ is denoted the coefficient of determination. In
our correlation analysis of functional parameters, we 
shall be inspecting matrices of $r_{AB}^2$.

\subsection{Numerical considerations}

In this work, pertaining to well-bound systems, we apply the HF+BCS approach rather than the full HFB; hence, the canonical wave functions are approximated by the HF orbitals and the related occupations are given by  the standard BCS amplitudes.
Spherical Hartree-Fock wave functions, densities, and mean-fields are
represented on a 1D grid in spherical coordinates \cite{Rei91aR}. We use a grid spacing of 0.3\,fm
and 32--48 grid points depending on system size. The HF+BCS
equations are solved with an accelerated gradient iteration technique
\cite{Rei82a}.

We use the soft pairing cutoff \cite{Kri90a} in BCS defined by the
cutoff energy $\epsilon_{\rm cut}=15$\,MeV with respect to the Fermi
energy and the width $\Delta\epsilon=\epsilon_{\rm cut}/10$.  Although
seemingly straightforward, pairing raises a subtle problem around a
phase transition between paired and normal state around closed
shells, which may result in numerical
instabilities. We avoid such unphysical behavior by using the
stabilized pairing of \cite{Erl08a} arranged to guarantee in smooth
manner a minimal gap of $\Delta=0.3$\,MeV, which is well below the
typical pairing gap of 1--2 MeV; hence, it has negligible influence on
nuclear bulk properties as binding energy and radii.

Odd-$A$ nuclei are treated in the standard uniform filling
approximation to blocking, in which a blocked nucleon is put with
equal probability in each of the degenerate magnetic sub-states
\cite{Fayans2000,Dug01}; hence, time-reversal symmetry is
conserved.  To
find the ground state, we carry out blocked calculations for all
shells near the Fermi energy and select the blocked state with lowest
energy. It is to be noted that the data on odd-$A$ nuclei contain
contributions from the  polarization effects due to the
odd nucleon. The experimental data in Table~\ref{fitobservables} on
3-point gaps $\DE$ and $\Dr$ were selected in such a way that the
impact of deformation-polarization, spin-polarization, and correlation effects is minimal. While these quantities can be affected by time-odd
polarizations, the corresponding corrections are expected to be small
\cite{Sch10,Pot10a}.

To find the optimal parameter set of parameters $\mathbf{p}_0$ we carry out multidimensional minimization of $\chi^2$.  Here we use  two iterative strategies: a  multi-dimensional method of determinants  as outlined in Ref.~\cite{Bev69aB}, and succession of
one-dimensional minimizations according to Powells method
\cite{Pre92aB}. The first method is much faster and it has the
great advantage to provide the full covariance matrix, which is
needed for covariance analysis and error estimates
\cite{Reinhard_2013,Dob14a,Erl14b,Haver17}. However, it easily gets stuck in
conflicting situations where one or a few data points constitute a large faction
of $\chi^2$, as this often happens when trying to accommodate isotopic
shifts. Here we switch to the Powell method which is comparably slow
but unerringly drives $\chi^2$ to a minimum. To acquire more
confidence that the minimum found is global, we restart iterations
several times by stochastically stirring up the model parameters
$\mathbf{p}$.

\section{Results}\label{results}

\begin{table*}[htb]
\caption{\label{tab:NMP}
NMP of  Fayans and Skyrme functionals used in this work.}
  \begin{ruledtabular}
\begin{tabular}{lcccccc}
 NMP  & FaNDF$^0$ & Fy(std) &  Fy($\Delta r$) & Fy($\Delta r,\Delta r^{\rm oe}$) & SV-min & Sk+PFy
\\   
 \hline \\[-8pt]  
$\rho_\mathrm{eq}$ (fm$^{-3}$) &   0.160 &0.163$\pm$0.002 &
 0.160$\pm$0.002 & 0.161$\pm$0.001
& 0.162$\pm$0.001 &0.163$\pm$0.001
\\
$E/A$ (MeV) & $-$16.00 & $-$16.10$\pm$0.05 & $-$16.11$\pm$0.04
& $-$16.12$\pm$0.03
& $-$15.91$\pm$0.04 &$-$15.94$\pm$0.03
\\
 $K$  (MeV) &   219 & 219$\pm$15 & 219$\pm$12
& 220$\pm$18
& 222$\pm$7 & 229$\pm$5
\\
 $J$ (MeV) &  30 & 31$\pm$2 & 29$\pm$2
 & 30$\pm$1
 & 31$\pm$2 & 30$\pm$1
\\
 $L$ (MeV)  &   30 & 59$\pm$22 & 30$\pm$24
&  35$\pm$21
&  45$\pm$26 & 18$\pm$18
\end{tabular}
\end{ruledtabular}
\end{table*} 
%%% 

\subsection{Nuclear Matter Parameters}

The volume term $ {\cal E}_{{\rm Fy}}^\mathrm{v}$ of Fayans EDF
determines its nuclear matter parameters. As a matter of fact, the
volume term coupling constants of the original FaNDF$^0$
parametrization \cite{Fayans1998} were fixed by fitting them to the
equation of state of symmetric infinite nuclear
matter. Table~\ref{tab:NMP} displays NMP of selected Fayans and Skyrme
functionals used and optimized in this work.  It is satisfying to see
that the values of NMP in Table\,\ref{tab:NMP} are consistent (within
error bars) with each other and with the range allowed by theory
\cite{Nazarewicz2014} and experiment/observations
\cite{Lattimer2014,Oertel17}.  The only notable exception is a
relatively low mean value of $L$ obtained in Sk+PFy. Considering that
Sk+PFy and SV-min are based on the same particle-hole interaction
parametrized to a very similar dataset, it is surprising to see an
unexpectedly large impact of the additional data on differential
charge radii on the symmetry energy slope. We checked that the Sk+PFy
fit without a constrain on isotopic shift yields $L=41$\,MeV, which is
a typical value as in other parametrizations. We also remark that $L$
is also sensitive to the density dependence of the pairing channel. By
re-optimizing Sk+PFy with the pairing functional having different
density dependence, $\gamma=2/3$, we obtain  $L=-3\pm 17$\,MeV, which is unreasonably low. As seen in Fig.~\ref{gammadependence}
below, such a low value of $\gamma$ is disfavored by our optimization
protocol.

%%%%%%
\begin{table}[htb]
\caption{\label{tab:params} Parameters of the various Fayans
  functionals optimized in this work compared to the original
  FaNDF$^0$ parametrization. The parameters fitted in Fy(std),
  Fy($\Delta r$), and Fy($\Delta r,\Delta r^{\rm oe}$) are displayed
  in the upper panel while the lower panel lists those that are
  fixed. The parameters are given with the number of digits as
  required for sufficient precision of the calculations.
  $\rho_{\mathrm{sat}}$ is in fm$^{-3}$; other parameters are
  dimensionless.}
  \begin{ruledtabular}
  \begin{tabular}{lcccc}
 & FaNDF$^0$ & Fy(std) & Fy($\Delta r$) & Fy($\Delta r,\Delta r^{\rm oe}$) \\
 \hline \\[-8pt]  
${a_+^\mathrm{v}}$  &  $-$9.559  & $-$9.495922&$-$9.542989 &$-$9.534242
\\
${h_{1+}^\mathrm{v}} $& 0.633  & 0.6271056&0.6323225 & 0.6317662
\\
${h_{2+}^\mathrm{v}} $& 0.131  & 0.1452334&0.1343786 & 0.1348917
\\
${a_-^\mathrm{v}}$ & 4.428  & 12.4741& 4.18236 & 3.96619
\\
${h_{1-}^\mathrm{v}}$ & 0.250  & $-$1.38754& 0.253776 &0.206941
\\
${h_{2-}^\mathrm{v}}$ & 1.300 & 19.3795&  1.21502 & 1.16421
\\
${a_+^\mathrm{s}} $& 0.600 & 0.5241615&0.6047266 & 0.5917059
\\
${h_{\nabla}^\mathrm{s}}$ & 0.440 & 0.0992&0.6656 &  0.4861
\\
${\kappa}$ & 0.19 & 0.189640 & 0.187922 & 0.197785
\\
${\kappa'}$ & 0.0 & 0.0250 & $-$0.0237  & $-$0.0052
\\
${f_\mathrm{ex}^\xi}$ &  $-$2.8& $-$1.636& $-$4.472 & $-$4.265
\\
${h_+^\xi}$   & 2.8 &  1.130& 4.229   &3.9618
\\
${h_\nabla^\xi}$& 2.2& 0.013 &  3.227  &3.8732
\\[4pt]
 \hline \\[-8pt]
${\sigma}$         & 1/3  & 1/3 & 1/3 & 1/3
\\
${h_{+}^\mathrm{s}}$ & ${h_{2+}^\mathrm{v}}$ & 0&0 & 0
\\
${h_\mathrm{Coul}}$ &  0.941 & 0 & 0 & 0
\\
${\rho_{\rm sat}}$  & 0.160  & 0.160 & 0.160& 0.160
\\
${\rho_{\mathrm{pair}}}$ & ${\rho_{\rm sat}}$ 
 &  ${\rho_{\rm sat}}$  &  ${\rho_{\rm sat}}$ &  ${\rho_{\rm sat}}$ 
\\
${\gamma}$  & 1 & 2/3 & 2/3 & 2/3
\end{tabular}
\end{ruledtabular}
\end{table}

\begin{table}[htb]
\caption{\label{tab:paramsSk}
Parameters of the Skyrme  functionals SV-min and Sky+PFy other than NMP shown in Table~\ref{tab:NMP}:
$m^*/m$ and $\kappa_\mathrm{TRK}$ are dimensionless; $C^{\rho\Delta\rho}_t$ and $C^{\rho\nabla J}_t$
are  in MeV\,fm$^5$; and $\rho_{\mathrm{pair}}$ is in fm$^{-3}$. The remaining  pairing parameters 
of Sky+PFy are: $\rho_{\mathrm{sat}}$=0.16\,fm$^{-3}$,  $\gamma=1$,
${f_\mathrm{ex}^\xi}=-3.3247\pm0.31$,
${h_+^\xi}= 2.7952\pm0.35$, and
${h_\nabla^\xi}=4.8432\pm0.70$.
}
  \begin{ruledtabular}
\begin{tabular}{lcc}
  & SV-min  & Sky+PFy
\\
 \hline \\[-8pt]  
 $m^*/m$ & 0.9518$\pm$0.07 & 1.0117$\pm$0.08
\\
$\kappa_\mathrm{\rm TRK}$ & 0.0765$\pm$0.28 & 0.1167$\pm$0.13
\\
$C^{\rho\Delta\rho}_0$& $-$89.205$\pm$5.2 & $-$23.499$\pm$6.5 \\
$C^{\rho\Delta\rho}_1$& $-$35.377$\pm$39 & 25.734$\pm$17 \\
$C^{\rho\nabla J}_0$& $-$101.581$\pm$4.9 & $-$72.390$\pm$3.9 \\
$C^{\rho\nabla J}_1$& $-$22.968$\pm$19 & 8.773$\pm$40 \\
$\rho_{\mathrm{pair}}$ & 0.211591$\pm$0.05 & 0.16
\end{tabular}
\end{ruledtabular}
\end{table}

\subsection{Optimized Functionals}

The parameters of EDFs used and optimized in this work are displayed in
Tables~\ref{tab:params} (Fayans functionals) and \ref{tab:paramsSk} (Skyrme).
The coupling constants characterizing the isoscalar part of the volume term
(\ref{EFay-dens}) are similar between different Fayans functionals. This is not surprising in light of similarity of NMP in Table~\ref{tab:NMP}.
The parameters of the particle-hole interaction of Fy($\Delta r$) and  Fy($\Delta r,\Delta r^{\rm oe}$)  are not far from the original FaNDF$^0$. This is not the case for the isovector-volume and surface terms of Fy(std). This suggests that constraining $\delta \langle r^2\rangle$ plays a key role for determining the spectroscopic quality of the Fayans model. Indeed, Table~\ref{tab:params}
shows that the coupling constants $h_{\nabla}^{\rm s}$ and ${h_\nabla^\xi}$ determining the strengths of the gradient terms $\propto(\vec{\nabla}\rho_0)^2$ in the Fayans surface and pairing functional, respectively,  are increased by orders of magnitude when the data on differential radii are added to the dataset. 
Another interesting outcome is the significant difference between the parameters of SV-min and Sk+PFy shown in Table~\ref{tab:paramsSk}, again primarily attributed to the 
additional data on differential radii used to constrain Sk+PFy. 
It is worth noting that the value of ${h_\nabla^\xi}$ of Sk+PFy is in the same range that that of Fy($\Delta r$).

\begin{figure}[htb]
\includegraphics[width=0.8\linewidth]{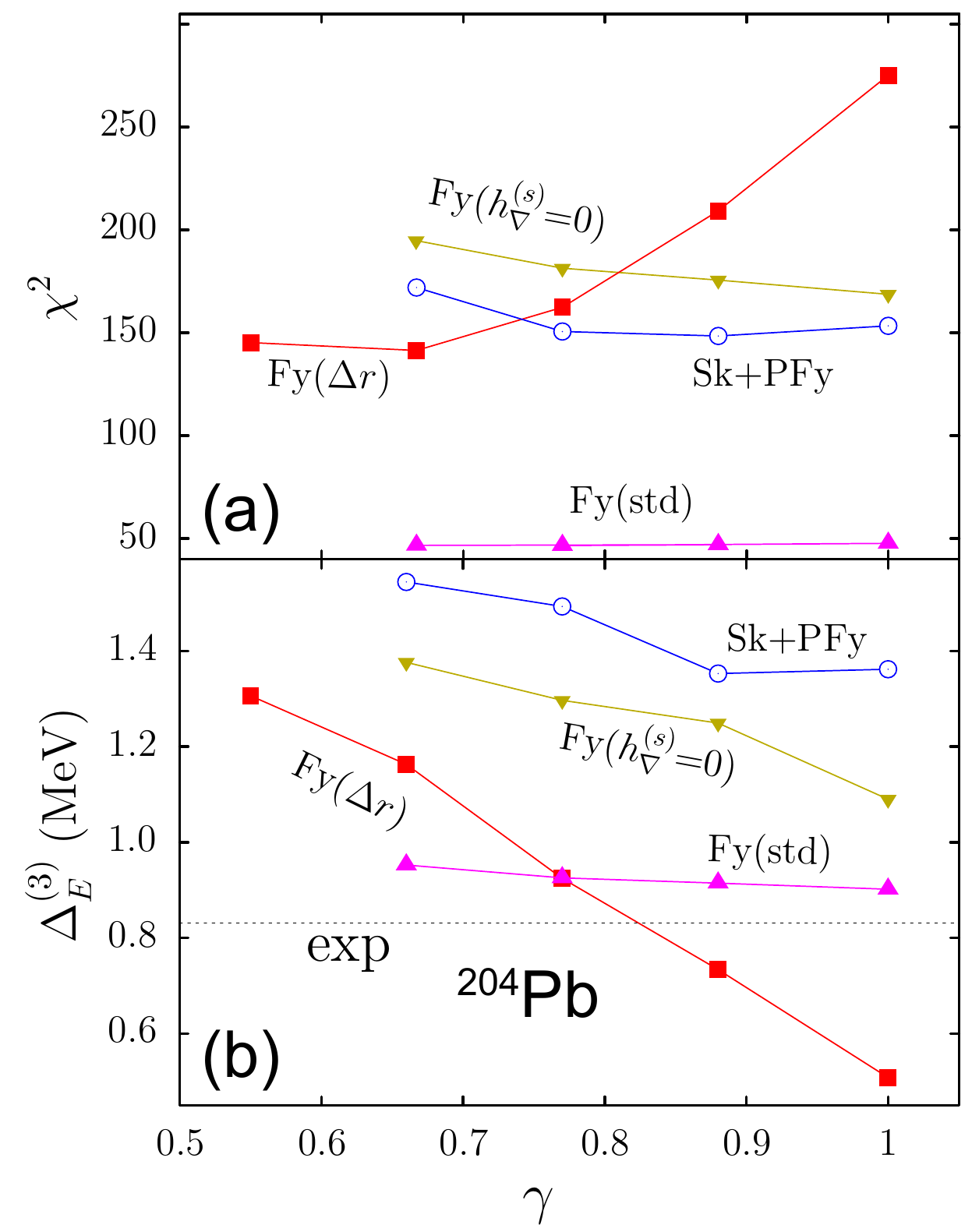}
\caption{\label{gammadependence}
Quality measure $\chi^2$ (top) and odd-even binding-energy staggering   $\DE$ for $^{204}$Pb (bottom) and  as functions of  the $\gamma$ parameter of the  Fayans pairing functional for Fy(std), Fy($\Delta r$), Fy(${h_\nabla^s}$), and Sk+PFy.}
\end{figure}

The parameter  $\gamma$ entering the density dependence  of the  Fayans pairing functional is of particular interest as it changes the character of pairing potential from surface-type ($\gamma<1, \rho_{\rm pair}\le\rho_{\rm sat}$) to mixed-type ($\gamma\ge 1, \rho_{\rm pair}> \rho_{\rm sat}$).
Figure~\ref{gammadependence}(a) shows the dependence of the quality measure (\ref{eq:chi2}) on $\gamma$.
%%%%
The parametrization  Fy(std) is fairly insensitive to $\gamma$ as the value of $h_+^\xi$ is significantly reduced for this parametrization as compared to the other ones in Fig.~\ref{gammadependence}. The constraint on differential radii in Fy($\Delta r$) clearly favors $\gamma=2/3$. This is due to the interplay between the surface-gradient effect and pairing term of the Fayans functional. Indeed, eliminating the surface-gradient correction in Fy(${h_\nabla^s}$=0) tends to favor larger values of $\gamma$, and this is in line with the Sk+PFy result. To illustrate the impact of $\gamma$ and the data on differential radii  on odd-even binding energy staggering, Fig.~\ref{gammadependence}(b) shows $\DE$ in $^{204}$Pb. The prediction of  Fy(std) is close to experiment and hardly varies with $\gamma$.
It is seen that the additional dataset  $\Delta r$ does impact pairing correlations significantly, and -- by increasing $h_+^\xi$  -- it gives rise to  large variations of $\DE$ with $\gamma$.

%%%%
\begin{figure*}[htb]
\includegraphics[width=0.6\linewidth]{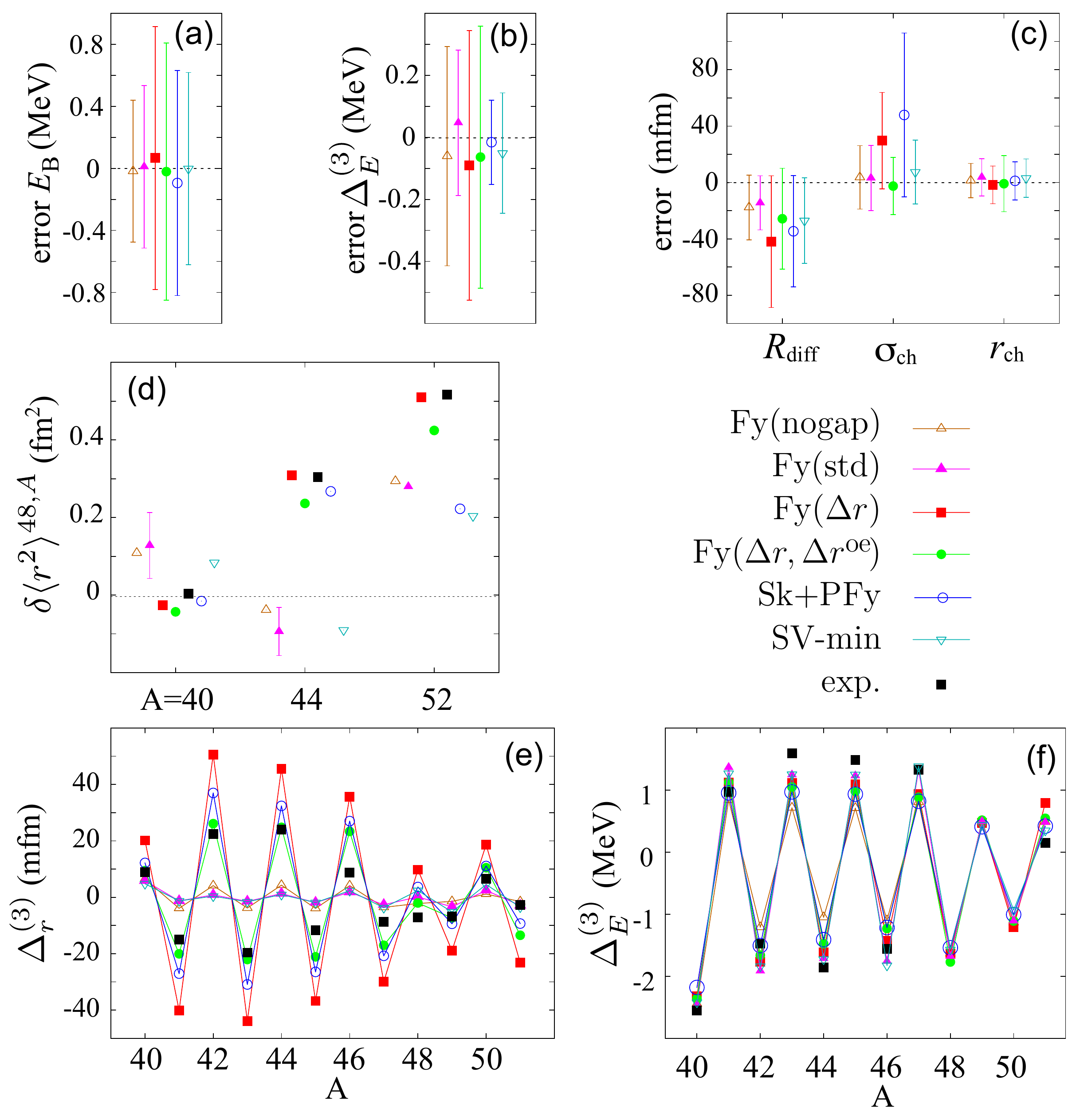}
\caption{\label{fig:staggering-sum}
 Summary of global performance of Fayans and Skyrme parametrizations used/optimized in this work; for naming convention see Sec.~\ref{variants}. Top: average and r.m.s. deviations  (\ref{eq:quali}) for  selected  observables of the fit.  Middle: Differential  charge radii
  in Ca.  Bottom: odd-even staggering of charge radii and binding
  energies in Ca isotopes. 
 }
\end{figure*}
%%%

\subsection{Global performance}

Figure~\ref{fig:staggering-sum} illustrates the performance of Fayans and Skyrme parametrizations used/optimized in this work. The optimization quality is measured in terms 
of average and r.m.s. deviations  of fit observables from experiment:
\begin{subequations}
\label{eq:quali}
\begin{eqnarray}
  \overline{\mathcal{D}}_\mathcal{O}
  &=&
  \frac{\sum_n \left(\mathcal{O}_n^\mathrm{th}-\mathcal{O}_n^\mathrm{exp}\right)}
       {N_\mathrm{data}},
\\
  \overline{\Delta\mathcal{D}}_\mathcal{O}
  &=&
  \sqrt{
  \frac{\sum_n\left(\mathcal{O}_n^\mathrm{th}
                -\mathcal{O_n}^\mathrm{exp}\right)^2}
       {N_\mathrm{data}}
  -
  \overline{\mathcal{D}}_\mathcal{O}^2
  },
\end{eqnarray}
\end{subequations}
where $\mathcal{O}$ stands for one of the observables: $E_{\rm B},  \DE$, and charge form factor characteristics: $R_\mathrm{diff}$, $\sigma_{\rm ch}$, and $r_\mathrm{ch}$.
The  parametrizations Fy(std), Fy(nogap),  and SV-min,  optimized to  similar  datasets yield results of comparable quality for bulk observables 
(energies and charge form factor  characteristics). This performance  deteriorates as the data on 
$\Delta r$ and $\Delta r^{\rm oe}$ are added to the dataset. This is seen in large error bars on the values of $E_{\rm B}$, $\DE$, $R_\mathrm{diff}$, $\sigma_{\rm ch}$, and $r_\mathrm{ch}$ predicted by
Fy($\Delta r$) and  Fy($\Delta r,\Delta r^{\rm oe}$).

Differential radii for the Ca isotopes are are shown in Fig.~\ref{fig:staggering-sum}(d). One can see that by constraining the functional by additional $\Delta r$  data, as done  for Fy($\Delta r$),  Fy($\Delta r,\Delta r^{\rm oe}$), and Sk+PFy, one is able to reproduce experiment for $A=40$ and 44. The unexpectedly  large charge radius in  $^{52}$Ca \cite{GarciaRuiz16} is reproduced by the $\Delta r$-constrained  Fayans functionals but it is underestimated in Sk+PFy and also in  Fy(std), Fy(nogap),  and SV-min.

Interestingly, the odd-even energy staggering $\DE$ in Ca is reproduced 
reasonably well by all models, see Figs.~\ref{fig:staggering-sum}(b)
and (f).  When looking into details, however, one can see that by adding data on $\DE$ to the basic dataset, helps reducing theoretical error when going from Fy(nogap) to Fy(std). 

As pointed out in Ref.~\cite{FayZ96,Fayans2000}, the odd-even staggering of charge radii can be attributed to the contribution to the mean-field potential arising from the pairing interaction (\ref{eq:ep2}):
\begin{subequations}\label{hpair}
\begin{eqnarray}
h_{\rm pair}  &=&  \sfrac{2{\varepsilon_F}}{3\rho_{\rm sat}}\left\{ 
h_+^\xi \gamma x_{0}^{\gamma-1}({\breve\rho_n}^2
+{\breve\rho_p}^2)\right. 
\label{hpair1}
\\ 
 && 
 \left.-2h_\nabla^\xi r_s^2 \vec{\nabla}\left[ ({\breve\rho_n}^2 +{\breve\rho_p}^2)\vec{\nabla}x_{0}\right]
\right\},  
\label{hpair2}
\end{eqnarray}
\end{subequations}
where we explicitly put $x_{\rm pair}=x_{0}$ (as $\rho_{\rm pair}=\rho_{\rm sat}$). The field
(\ref{hpair}) produces a direct coupling between the isoscalar particle density $\rho_0$ and  pairing densities $\breve\rho_n$ and $\breve\rho_p$, which results in the  odd-even staggering in charge radii. Indeed, the  blocking effect in an odd-$A$ nucleus yields  a reduced pairing  density $\breve\rho_n$, which in turn impacts the proton (or charge) density, hence  $r_{\rm ch}$. 
Figure~\ref{fig:staggering-sum}(e) illustrates the importance of the couplings (\ref{hpair}) for the Ca chain. The  parametrizations Fy(std), Fy(nogap),  and SV-min
dramatically underestimate the magnitude of $\Dr$. Indeed, in these models the coupling constant  $h_\nabla^\xi$ is either zero (SV-min) or very small (Fy(std) and  Fy(nogap)), and $\Dr$ is  driven by the first term (\ref{hpair1}). In other models, having large values of $h_\nabla^\xi$, the staggering primarily results from  the second term  (\ref{hpair2}).
Here we see that the additional information contained in  datasets  $\Delta r$ and $\Delta r^{\rm oe}$ is absolutely crucial for boosting $h_\nabla^\xi$.

To check the importance of  c.m. correction, we carried out two sets of calculations  for the observables shown in  Fig.~\ref{fig:staggering-sum}:  with and without c.m. correction. We conclude that the impact
of the c.m. term is negligible for radial properties, and plays fairly minor role for energies. In the following, we stick to calculations with the c.m. term added, but its effect is should be viewed as secondary.

\begin{figure}[htb]
\centerline{\includegraphics[width=\linewidth]{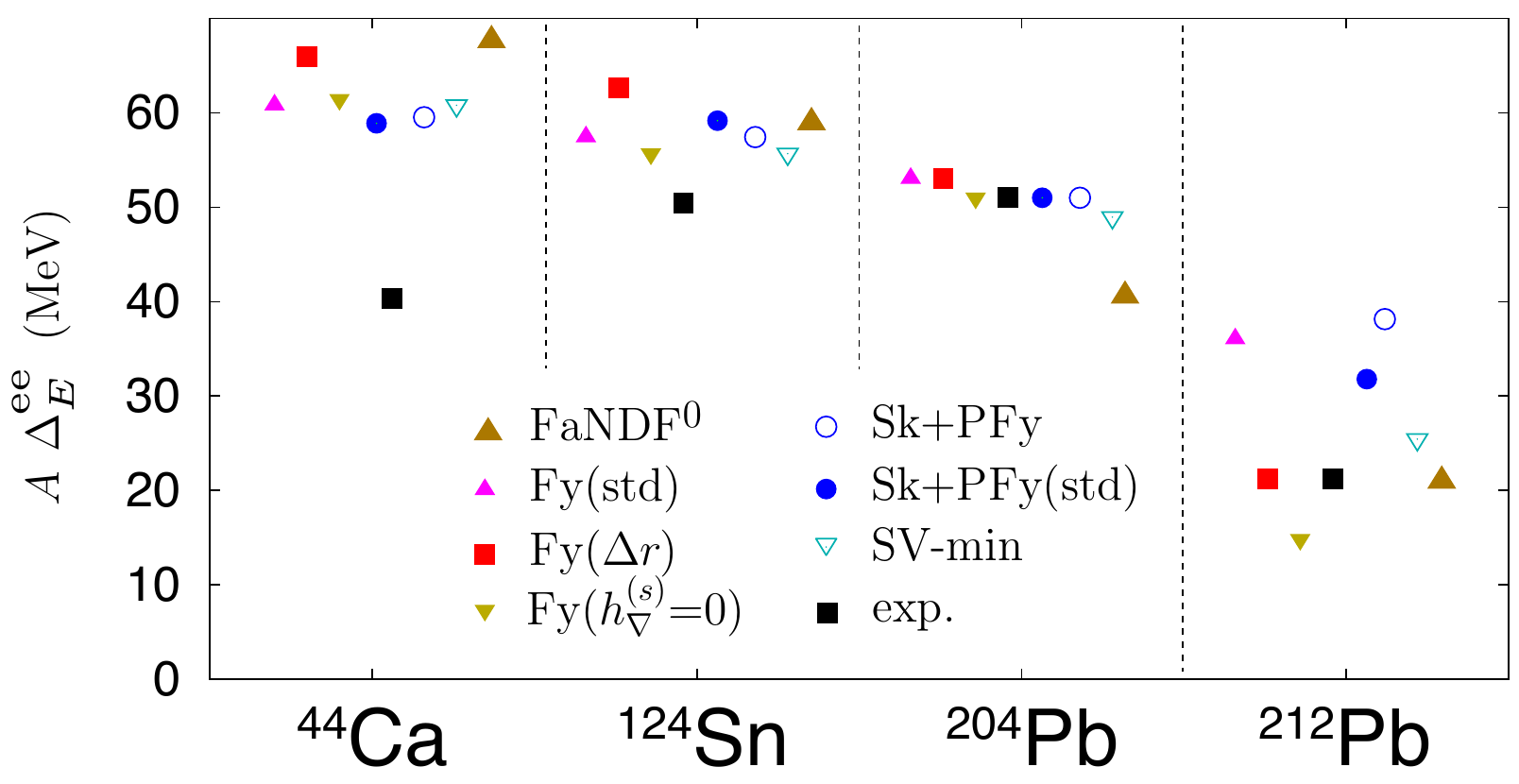}}
\caption{\label{pairingrot} 
Three-point binding energy difference $\De$ (multiplied by the mass number to easily compare different systems)   for the spherical open-shell nuclei  $^{44}$Ca,
$^{124}$Sn, $^{204}$Pb, and $^{212}$Pb computed with  selected Fayans and Skyrme functionals and compared to experiment. 
 }
\end{figure}
%%%%
The performance of optimized functionals with respect to $\De$ in $^{44}$Ca,
$^{124}$Sn, $^{204}$Pb, and $^{212}$Pb is illustrated in Fig.~\ref{pairingrot}. 
Since the Fayans pairing functional  exhibits an $A$-dependent scaling,
in order to include FaNDF$^0$ results in the mix, we scaled the parameters of FaNDF$^0$  pairing functional given in Table~\ref{tab:params}   by an overall factor 1.2, which yields reasonable results for all nuclei in the sample considered.
It is seen that for  $^{44}$Ca,
$^{124}$Sn, and $^{204}$Pb, there is a great consistency between different functionals employed. Indeed,  most information on pairing correlations is contained in the mass surface of even-even nuclei \cite{Hinohara16}, and all  our functionals shown in Fig.~\ref{pairingrot} utilize masses from  the basic dataset of SV-min. The discrepancy between calculated and experimental values of $\De$ was discussed in Ref.~\cite{Hinohara16}: the static pairing in $^{42,46}$Ca is close to the unphysical pairing phase transition. For $^{212}$Pb, the spread of predicted values of $\De$ is significant, with Fy(std) and  Sk+PFy functionals overestimating pairing correlations significantly.

%%%%%%%%%%
\begin{figure}[htb]
\includegraphics[width=1.0\linewidth]{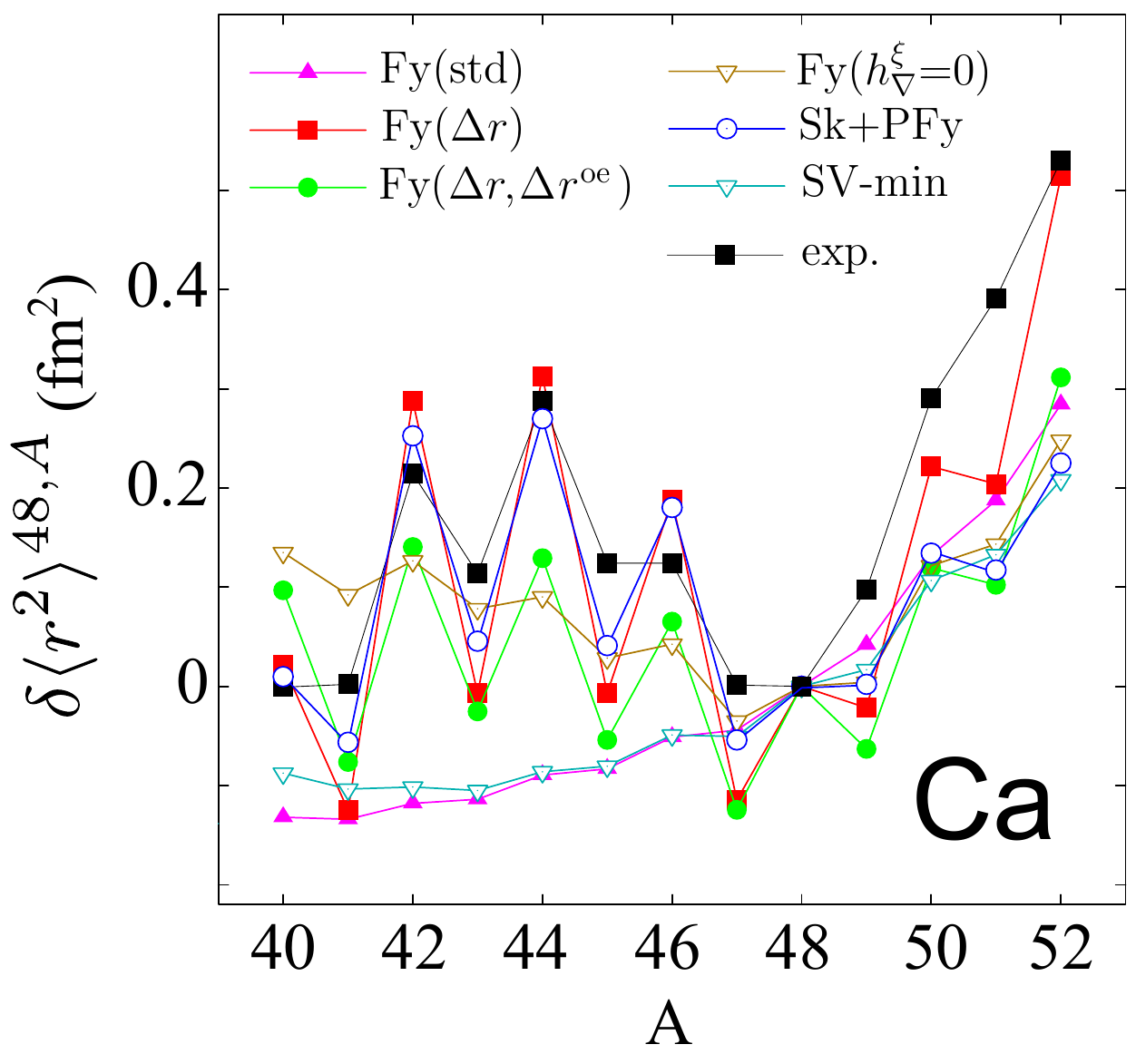}
\caption{\label{fig:trendisor-med} Differential charge radii $\delta
 \langle  r^2\rangle^{48,A}$ (\ref{rdiff}) in Ca predicted by 
Fayans and Skyrme parametrizations as indicated, and compared to experiment.}
\end{figure}
%%%%%%%%%
Figure \ref{fig:trendisor-med} shows predicted differential radii 
$\delta \langle  r^2\rangle^{48,A}$ for
the Ca chain.  The functionals optimized to the basic dataset, Fy(std) and SV-min,
exhibit characteristic monotonic dependence on $A$ \cite{Minamisono16}, and the odd-even affect is almost nonexistent. As discussed earlier, the latter effect can be attributed to  $h_\nabla^\xi$.
The Fayans parametrization
without the pairing gradient term Fy(${h}_\nabla^\xi$=0) does not perform
well either; while the magnitude of the odd-even staggering is slightly increased, the overall trend between $^{40}$Ca and $^{48}$Ca is incorrect.
It is only in the  functionals with the full Fayans pairing and
constraints on differential radii that large $\Dr$ values can be
obtained. 

Another set of radius differences, which had raised much attention in the past is that along the Pb chain. It has been argued in \cite{Fayans2000,Tolokonnikov2010} that the Fayans pairing functional allows to reproduce the  kink in $\Drr$ at $^{208}$Pb. In this case, however, pairing seems not to be the only influential agent. For instance, relativistic and Skyrme mean-field models can associate the kink with spin-orbit coupling \cite{Sha93b,Rei95a}. As it is a task of its own to disentangle  various influences on the kink in $^{208}$Pb, we are not addressing this observable here. 
 
The functionals with the full Fayans pairing, namely Fy($\Delta r$),
 Fy($\Delta r,\Delta r^{\rm oe}$), and Sk+PFy, perform well up to $^{48}$Ca. However, the $\delta \langle  r^2\rangle$ values in $^{49-52}$Ca are  underestimated in all models except for Fy($\Delta r$). This seems to suggest that correlations beyond mean field \cite{GarciaRuiz16,Minamisono16,Saperstein2016} can play a role there.
It is also interesting to note that  the attempt to
tune $\Dr$ in Fy($\Delta r,\Delta r^{\rm oe}$) results in a deterioration of  the reproduction of  differential radii $\Delta r$.

\begin{figure}[htb]
\includegraphics[width=0.8\linewidth]{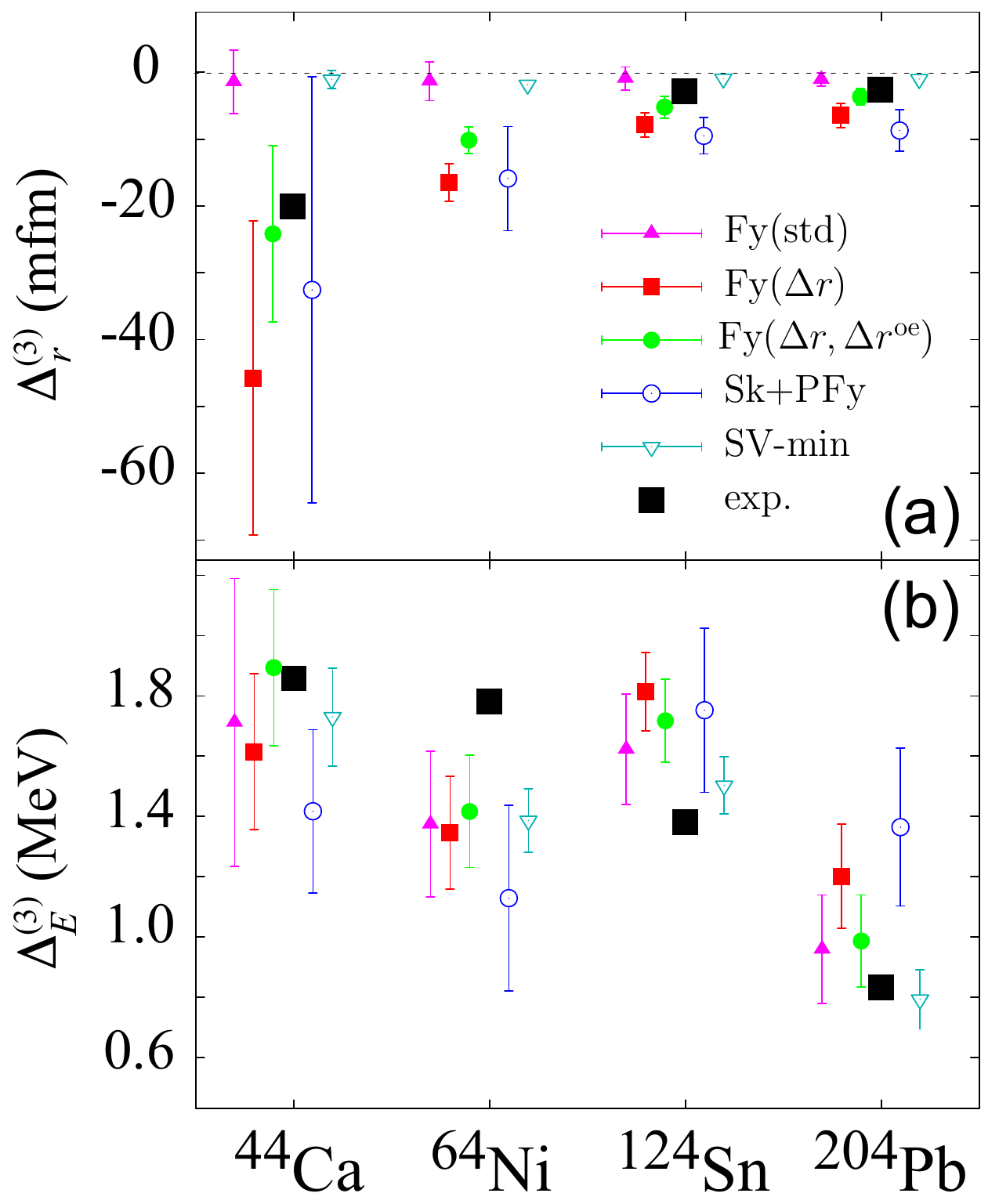}
\caption{\label{fig:values-3gap-med} Predicted mean values and variances of
  three-point neutron differences $\Dr$ (top) and $\DE$ (bottom)
  in $^{44}$Ca, $^{64}$Ni, $^{124}$Sn, and $^{204}$Pb, predicted by selected Fayans and Skyrme functionals, and compared to experiment.}
\end{figure}

\begin{figure}[htb]
\centerline{\includegraphics[width=\linewidth]{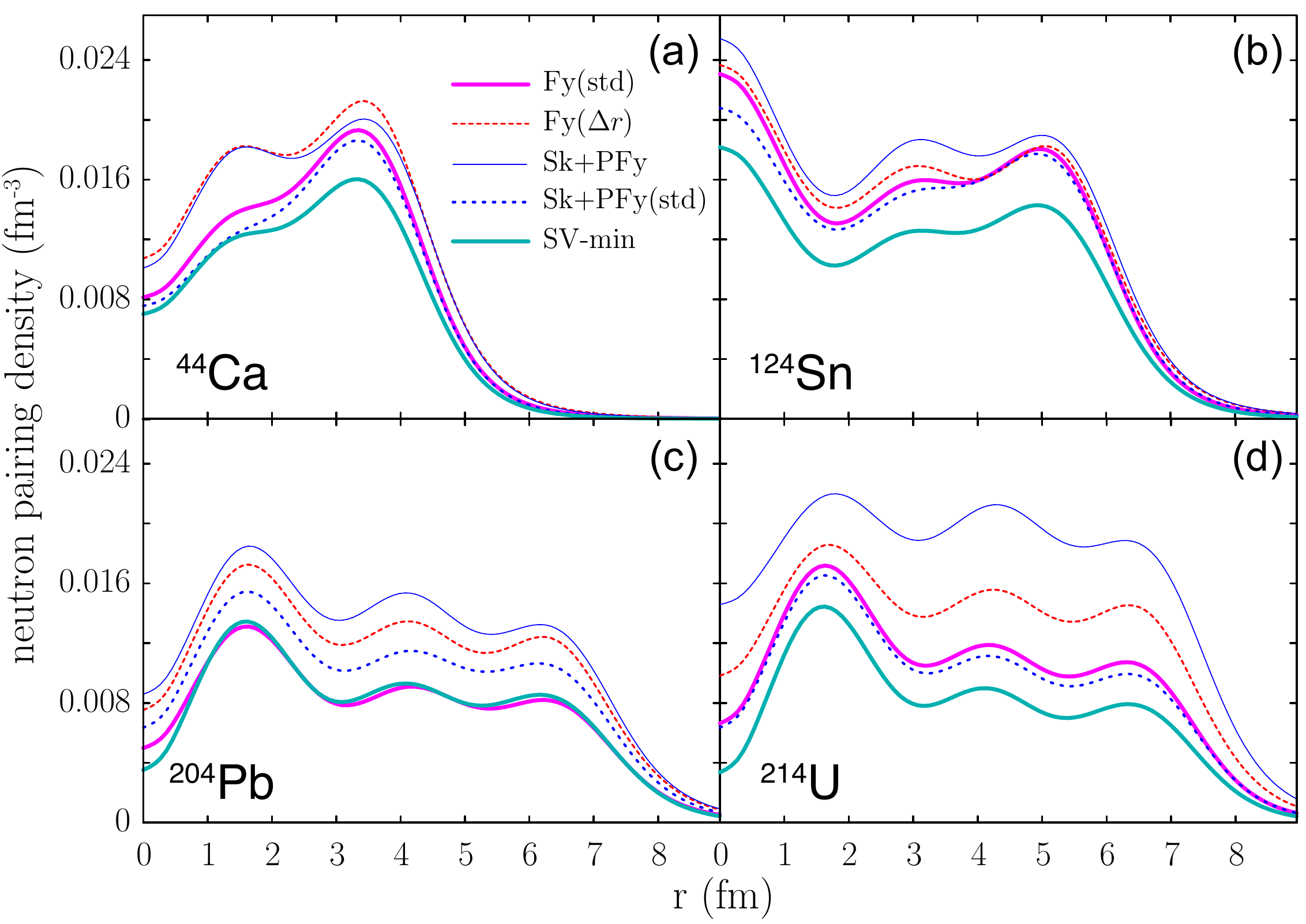}}
\caption{\label{pairdens} Neutron pairing densities ${\breve\rho_n}$   in  spherical configurations of $^{44}$Ca, $^{124}$Sn, $^{204}$Pb, and $^{214}$U, predicted by selected Fayans and Skyrme functionals.
 }
\end{figure}

To illustrate the performance of our optimized functionals for odd-even staggering  in medium-mass and heavy nuclei, shown in Fig.~\ref{fig:values-3gap-med} are their predictions for neutron values of  $\Dr$ and $\DE$ in $^{44}$Ca, $^{64}$Ni, $^{124}$Sn, and $^{204}$Pb. In accordance with the  discussion around Fig.~\ref{fig:trendisor-med}, Fy(std) and SV-min fail in 
reproducing $\Dr$, and Fy($\Delta r,\Delta r^{\rm oe}$) and Sk+PFy have a comparable performance.
The results for $\DE$ in  $^{44}$Ca, $^{64}$Ni, and $^{124}$Sn are consistent across different functionals. A large difference between experiment and Fy($\Delta r$) and Sk+PFy results  for $^{204}$Pb has been puzzling. To explain this dramatic enhancement of pairing predicted by these models, in Fig.~\ref{pairdens} we show the neutron pairing density ${\breve\rho_n}$   in $^{44}$Ca, $^{124}$Sn, $^{204}$Pb, and $^{214}$U obtained in different parametrizations. ($^{214}$U is slightly deformed in its ground state. However, for the purpose of  this discussion, we considered  it spherical.) In general, the shape of ${\breve\rho_n}$ is predicted consistently by all functionals considered. What is different is the overall magnitude of pairing density. In particular, in heavy nuclei such as $^{204}$Pb and  $^{214}$U, ${\breve\rho_n}$ predicted by Sk+PFy and Fy($\Delta r$) becomes very large, and this results in unreasonably large pairing fields. This is indicative of somehow uncontrolled $A$-dependence of Fayans pairing functional and explains previous results of Ref.~\cite{Fayans2000}, where it was necessary to increase the strength of pairing functional by  as much as 35\% when going from Pb to Ca. 
%%%%

\begin{figure}[htb]
\centerline{\includegraphics[width=0.6\linewidth]{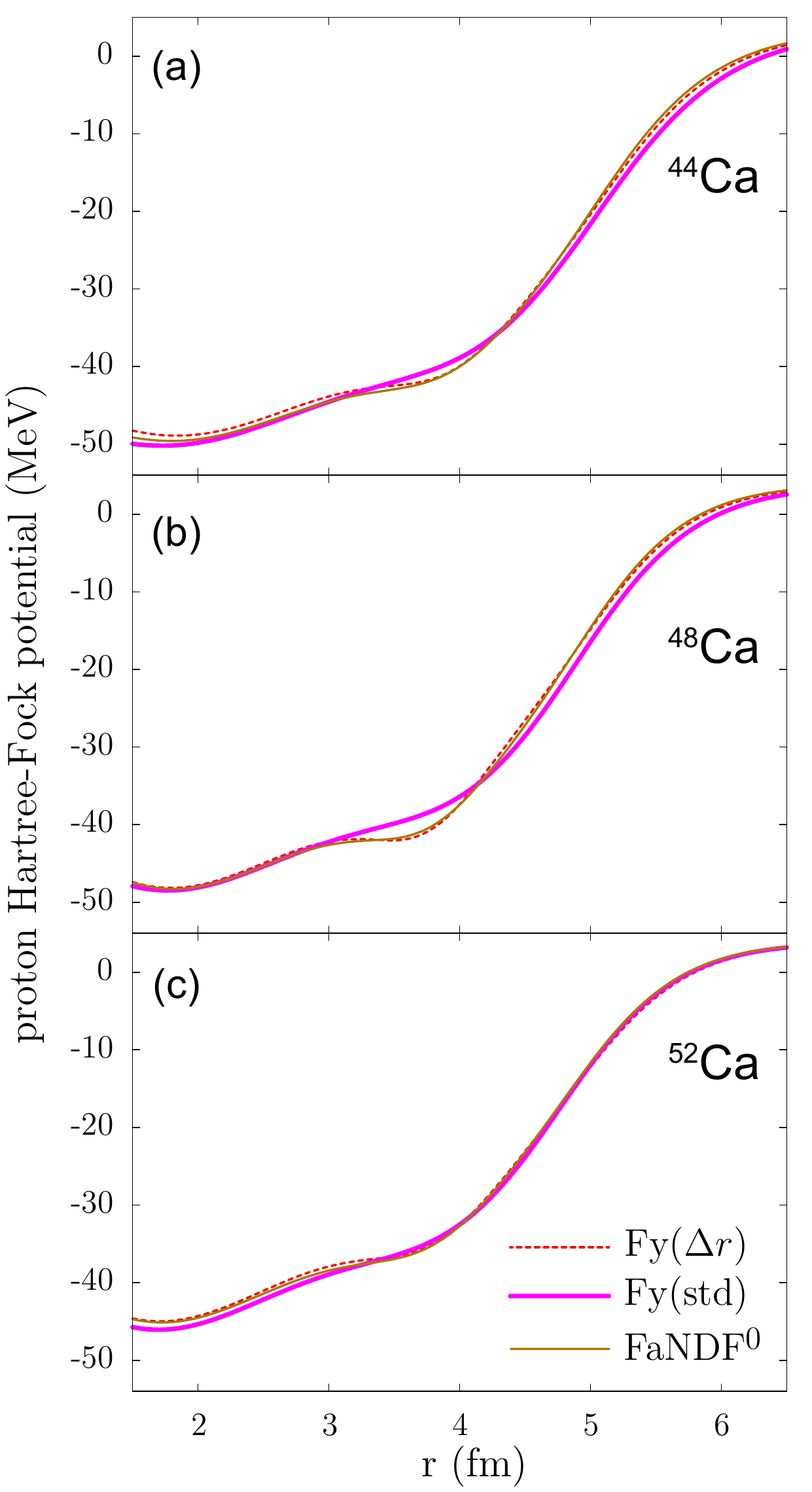}}
\caption{\label{fig:densCa-med} Proton HF potential for $^{44}$Ca (top), $^{48}$Ca (middle)
and $^{52}$Ca (bottom)
obtained in  Fy(std), Fy($\Delta r$), and FaNDF$^0$.
 }
\end{figure}
\begin{figure}[htb]
\centerline{\includegraphics[width=0.6\linewidth]{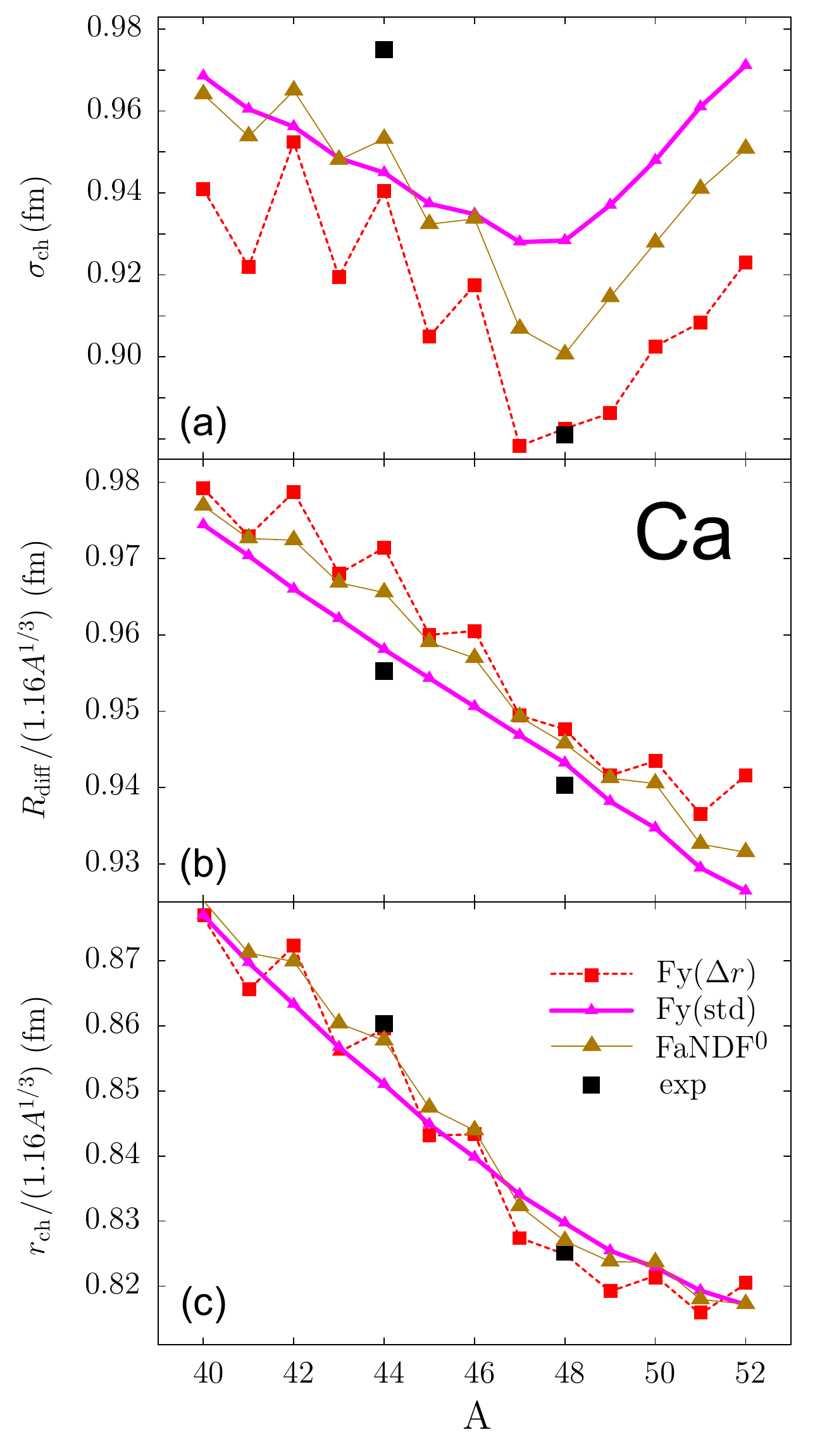}}
\caption{\label{fig:trendrms-med} Charge form factor characteristics,  $\sigma_{\rm ch}$,  $R_{\rm diff}$, and $r_{\rm ch}$,
 along the chain of Ca isotopes for  Fy(std), Fy($\Delta r$), and FaNDF$^0$. Experimental data are taken from Ref.~\cite{Fri82a}.
 }
\end{figure}

Figure \ref{fig:densCa-med} compares proton HF potentials of Fy(std) and Fy($\Delta r$) for
$^{44}$Ca and $^{48}$Ca. Both FaNDF$^0$ and Fy($\Delta r$) produce proton
potentials with pronounced flattening, or even small pockets,
in the surface region. Compared to Fy(std), this feature can be attributed to the large  parameters $h_\nabla^{\rm s}$ and $h_\nabla^\xi$, which define the strength of gradient terms. It is instructive to see how these terms influence surface properties of the charge form factor. To this end, in Fig.~\ref{fig:trendrms-med} we show the isotopic trends of charge
radii, diffraction radii,  and surface thickness predicted with Fy(std),  Fy($\Delta r$), and FaNDF$^0$ along the Ca chain.  These three quantities are related via \cite{Helm56,Fri82a,Mizutori00}
\begin{equation}\label{Helm}
r_{\rm ch}\approx \sqrt{3\over 5} \sqrt{R^2_{\rm diff}+5\sigma_{\rm ch}^2}.
\end{equation}
The impact of the additional data on differential radii  in Fy($\Delta r$)   is  significant;   the presence of this additional constraint results in a considerable reduction of
$\sigma_{\rm ch}$ and a simultaneous increase of $R_{\rm diff}$. As a consequence, the charge radius of $^{44}$Ca is increased, and that  of $^{48}$Ca is reduced, with respect to the Fy(std) prediction. While the properties of the charge form factor  of $^{48}$Ca  are well reproduced by Fy($\Delta r$), this is not the case for of $^{44}$Ca, where a good agreement for $r_{\rm ch}$ is obtained at the cost of underestimating $\sigma_{\rm ch}$ and overestimating of $R_{\rm diff}$. Another interesting lesson offered by Fig.~\ref{fig:trendrms-med} is that the odd-even staggering of charge radii in Fy($\Delta r$) and FaNDF$^0$ comes from  an appreciable odd-even effect in  $\sigma_{\rm ch}$ and  $R_{\rm diff}$: both quantities are reduced  in odd-$A$ isotopes as compared to their even-even neighbors. This effect is virtually nonexistent in Fy(std), which again highlights the impact  of  large density gradient terms in Fy($\Delta r$).

%%%

%%%
\begin{figure}[htb]
\includegraphics[width=\linewidth]{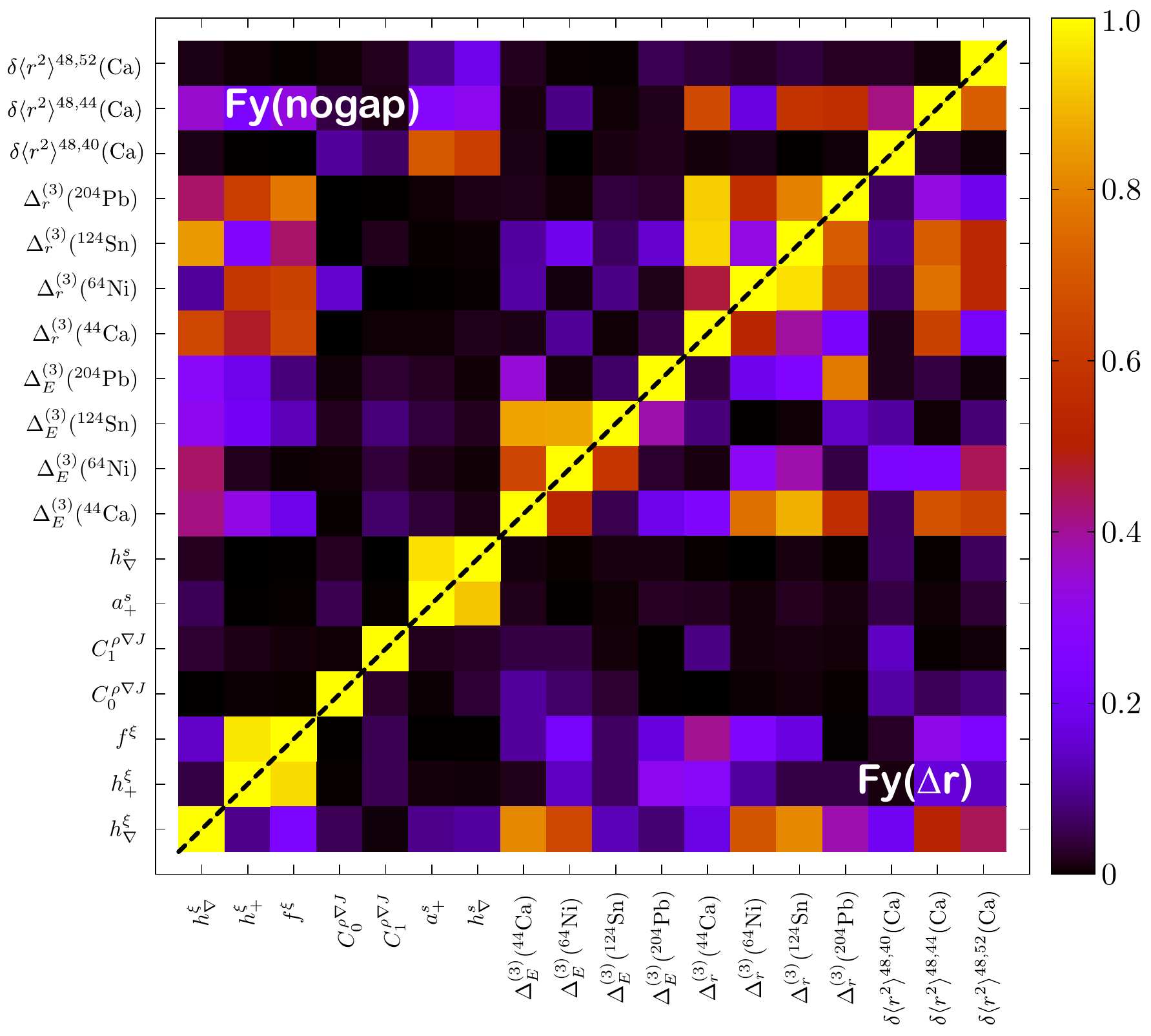}
\caption{\label{alignmmatrixFy} 
Matrices of coefficients of determination $r^2_{AB}$
for a selection of observables  computed with the Fayans functionals Fy(nogap) (upper triangle) and 
Fy($\Delta r$) (lower triangle).}
\end{figure}
%%%

\subsection{Correlations}

To understand better the impact of individual parameters of the Fayans functional on calculated observables, in Fig.~\ref{alignmmatrixFy} we show the coefficients of determination $r^2_{AB}$  for Fy(nogap) and  Fy($\Delta r$).
Since Fy(nogap) has not been constrained to differential radii, it is particularly suitable  for the analysis  of correlations related to $\Dr$ and $\Drr$.

That the  odd-even staggering in charge radii is primarily driven by the pairing functional (\ref{eq:ep2}) is shown by the large values of $r^2_{AB}$ between $\Dr$ and $f_\mathrm{ex}^\xi, h_+^\xi$, and $h_\nabla^\xi$.
The key value of $\Drr^{48,40}$(Ca) is primarily determined by the surface-energy coupling constants
$a_+^{\rm s}$ and $h_{\nabla}^{\rm s}$.
For  Fy($\Delta r$),   correlations between $\Drr$ values  and surface-energy parameters are gone, as the differential radii in Ca we constrained in the fit. Adding the dataset $\Delta r$ increases the impact of the pairing gradient term $h_\nabla^\xi$ on charge radii and $\DE$. For both functionals,  spin-orbit parameters $C_t^{\rho\nabla J}$ correlate with neither charge radii nor pairing.

\begin{figure}[htb]
\includegraphics[width=\linewidth]{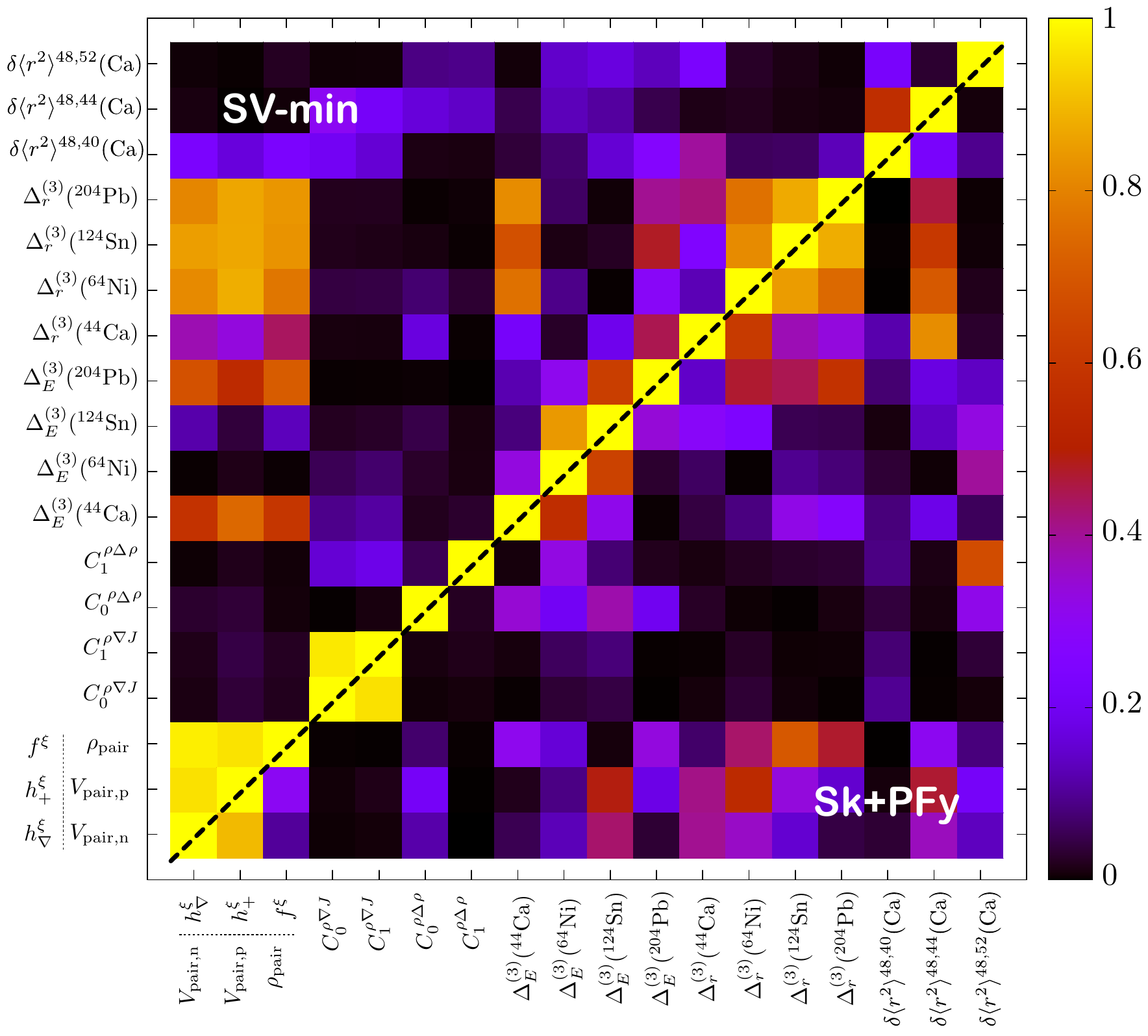}
\caption{\label{alignmmatrixSk} 
Similar as in  Fig.~\ref{alignmmatrixFy} except  for the Skyrme parametrizations
SV-min  (upper triangle) and Sk-PFy (lower triangle).}
\end{figure}

The matrices of $r_{AB}^2$ for the Skyrme functionals SV-min and
Sk-PFy are displayed in Fig.~\ref{alignmmatrixSk}. As SV-min and
Fy(nogap), and Sk+PFy and Fy($\Delta r$) have been optimized to the
same respective datasets, it is instructive to compare the
corresponding coefficients of determination.  In general, there is a
good correspondence.  In particular, for both Skyrme functionals, the
correlations of $\Drr$, $\Dr$, and $\DE$ with pairing parameters are
significant,  this even more so as they employ different pairing
  models. Interestingly, we find marginal correlations between
differential radii and parameters $C_t^{\rho\Delta\rho}$,
$C_t^{\rho\nabla J}$, and $C_t^{\rho\tau}$ (not shown). This suggests
that the impact of the surface energy on $\Drr$ is less pronounced in
the Skyrme case.
%%%%

\section{Conclusions}\label{conclusions}

By using the  tools of numerical optimization and linear regression, we studied 
properties of the  Fayans energy density functional, which is known to provide superb description of selected nuclear properties, such as charge radii and separation energies, and nuclear matter. By carefully selecting datasets of experimental data used in optimization, we generated functionals aimed to probe questions pertaining to different observables. The main conclusions of our study can be summarized as follows.

The pairing gradient term, controlled by the coupling constant $h_\nabla^\xi$, is arguably the most important ingredient in the Fayans functional. Without this term, it is impossible to reproduce the odd-even staggering $\Dr$ of charge radii and reproduce the intricate pattern of $r_{\rm ch}$ in the Ca chain. We note that by adding the Fayans pairing functional to the standard Skyrme functional, the resulting parametrization  Sk+PFy,
provides a comparable reproduction of the data  as Fy($\Delta r$). The large value of $\Drr^{52,48}$(Ca) \cite{GarciaRuiz16}, while  explained by Fy($\Delta r$), still remains a puzzle as other Fayans functionals and Sk+PFy significantly underestimate experimental value.

The surface gradient term controlled by  the coupling constant  $h_{\nabla}^{\rm s}$  is less influential. While it is driving the near-zero value of $\Drr^{40,48}$(Ca), it does not seem to be crucial  for other radius differences. Moreover, the gradient terms  of the Fayans model significantly change the surface behavior of the proton potential, as seen in Fig.~\ref{fig:densCa-med} for FaNDF$^0$ and Fy($\Delta r$).

The data   on $\Drr$   and $\Dr$  are important for characterizing the pairing functional.  
Our study demonstrates  that the coupling constants $h_{\nabla}^{\rm s}$ and ${h_\nabla^\xi}$ determining the strengths of the gradient terms are increased by orders of magnitude when the data on differential radii are added to the pool
of fit-observables. 
In particular, the correlation analysis  indicates that the  radius staggering  $\Dr$ is more sensitive to pairing than the 
energy staggering $\DE$.  Since the nuclear pairing functional is rather poorly constrained by experimental binding-energy differences alone, this has practically eliminated more sophisticated models of pairing EDF \cite{Per04,Yam12}. In this respect, new-quality data on differential radii can help  to calibrate   a properly generalized pairing functional.

By comparing the Fy(std) results with those obtained with with the
Fayans functional optimized to the basic+$\Delta E^{\rm ee}$ dataset
(employing even-even energy differences instead of odd-even
staggering), we find that the data on even-even binding energy
  differences $\De$ in open-shell nuclei carry very similar
  information content with respect to pairing as $\DE$, while the
  interpretation of $\De$ in terms of pairing correlations is more
  straightforward.  It is thus recommended that selected data on
  $\Drr$, $\Dr$, and $\De$ are used in the future functional
  optimizations.

The analysis presented in this paper should be viewed as a
useful starting point for future investigations. While the Fayans pairing functional has several attractive features, especially in the context of odd-even effect on charge radii, it fails short to reach the global description of pairing effects across the nuclear landscape. In particular, the parameters $f_\mathrm{ex}^\xi, h_+^\xi$, and $h_\nabla^\xi$, which provide excellent description   of
charge radii and pairing in spherical nuclei up to tin, dramatically overestimate pairing correlations around lead and in the actinides. This is consistent with the findings of Ref.~\cite{Fayans2000}, where an ad-hoc renormalization of pairing functional was imposed to provide agreement with experiment. Consequently,  extensions of the current Fayans pairing functional should be investigated. The current Fayans pairing functional is manifestly isoscalar. However, as the Coulomb term does affect pairing, many Skyrme parametrizations such as SV-min employ isovector pairing functionals,
see Ref.~\cite{Bertsch09} and references quoted therein. The   definition (\ref{eq:ep2}) can easily be extended to accommodate the isovector dependence. 

Extensions of the Fayans surface term (\ref{EFay-grad}) are also possible. By enlarging the experimental dataset, the parameter $h^{\rm s}_+$ could perhaps be pinned down with an acceptable accuracy. Moreover,   the  dependence of
${\cal E}_\mathrm{Fy}^\mathrm{s}$ on the isovector density $x_1$ should be considered, as the current parametrization cannot account for the surface-symmetry effects. Another strategy worth exploring is to replace the surface term (\ref{EFay-grad}) by the folded  DF3 expression \cite{Fay94,Kromer95,Horen96}.

The  extension of the Fayans functional to deformed nuclei \cite{Tolok15}, augmented by Fayans pairing, will allow the global optimization  of the extended Fayans  and Sk+PFy  models to diverse data on spherical and deformed nuclei, at the full deformed HFB level, using the well-tested UNEDF methodology \cite{Kortelainen_2010,Kor12a,Kor14a}. Work along these lines is in progress. 

\begin{acknowledgments}
We thank E.E. Saperstein and S.V. Tolokonnikov   for their invaluable help in
explaining  the subtilities of the Fayans functional.
This material is
based upon work supported by the U.S.\ Department of Energy, Office of
Science, Office of Nuclear Physics under  award numbers 
DE-SC0013365 (Michigan State University) and   DE-SC0008511 (NUCLEI
SciDAC-3 collaboration).
\end{acknowledgments}

\bibliographystyle{apsrev4-1}
\bibliography{Fy-paper}

%Merlin.mbs v4.21 2009-07-09.
\begin{thebibliography}{10}%
\makeatletter
\providecommand \@ifxundefined [1]{%
 \ifx #1\undefined \expandafter \@firstoftwo
 \else \expandafter \@secondoftwo
\fi
}%
\providecommand \@ifnum [1]{%
 \ifnum #1\expandafter \@firstoftwo
 \else \expandafter \@secondoftwo
\fi
}%
\providecommand \enquote [1]{``#1''}%
\providecommand \bibnamefont  [1]{#1}%
\providecommand \bibfnamefont [1]{#1}%
\providecommand \citenamefont [1]{#1}%
\providecommand\href[0]{\@sanitize\@href}%
\providecommand\@href[1]{\endgroup\@@startlink{#1}\endgroup\@@href}%
\providecommand\@@href[1]{#1\@@endlink}%
\providecommand \@sanitize [0]{\begingroup\catcode`\&12\catcode`\#12\relax}%
\@ifxundefined \pdfoutput {\@firstoftwo}{%
 \@ifnum{\z@=\pdfoutput}{\@firstoftwo}{\@secondoftwo}%
}{%
 \providecommand\@@startlink[1]{\leavevmode\special{html:<a href="#1">}}%
 \providecommand\@@endlink[0]{\special{html:</a>}}%
}{%
 \providecommand\@@startlink[1]{%
  \leavevmode
  \pdfstartlink
   attr{/Border[0 0 1 ]/H/I/C[0 1 1]}%
   user{/Subtype/Link/A<</Type/Action/S/URI/URI(#1)>>}%
  \relax
 }%
 \providecommand\@@endlink[0]{\pdfendlink}%
}%
\providecommand \url  [0]{\begingroup\@sanitize \@url }%
\providecommand \@url [1]{\endgroup\@href {#1}{\urlprefix}}%
\providecommand \urlprefix [0]{URL }%
\providecommand \Eprint[0]{\href }%
\@ifxundefined \urlstyle {%
  \providecommand \doi [1]{doi:\discretionary{}{}{}#1}%
}{%
  \providecommand \doi [0]{doi:\discretionary{}{}{}\begingroup
  \urlstyle{rm}\Url }%
}%
\providecommand \doibase [0]{http://dx.doi.org/}%
\providecommand \Doi[1]{\href{\doibase#1}}%
\providecommand \bibAnnote [3]{%
  \BibitemShut{#1}%
  \begin{quotation}\noindent
    \textsc{Key:}\ #2\\\textsc{Annotation:}\ #3%
  \end{quotation}%
}%
\providecommand \bibAnnoteFile [2]{%
  \IfFileExists{#2}{\bibAnnote {#1} {#2} {\input{#2}}}{}%
}%
\providecommand \typeout [0]{\immediate \write \m@ne }%
\providecommand \selectlanguage [0]{\@gobble}%
\providecommand \bibinfo [0]{\@secondoftwo}%
\providecommand \bibfield [0]{\@secondoftwo}%
\providecommand \translation [1]{[#1]}%
\providecommand \BibitemOpen[0]{}%
\providecommand \bibitemStop [0]{}%
\providecommand \bibitemNoStop [0]{.\EOS\space}%
\providecommand \EOS [0]{\spacefactor3000\relax}%
\providecommand \BibitemShut [1]{\csname bibitem#1\endcsname}%
%</preamble>
\bibitem{Hagen16}%
  \BibitemOpen
  \bibfield{author}{%
  \bibinfo {author} {\bibfnamefont{G.}~\bibnamefont{Hagen}}, \bibinfo {author}
  {\bibfnamefont{A.}~\bibnamefont{Ekstr\"om}}, \bibinfo {author}
  {\bibfnamefont{C.}~\bibnamefont{Forss\'en}}, \bibinfo {author}
  {\bibfnamefont{G.~R.}\ \bibnamefont{Jansen}}, \bibinfo {author}
  {\bibfnamefont{W.}~\bibnamefont{Nazarewicz}}, \bibinfo {author}
  {\bibfnamefont{T.}~\bibnamefont{Papenbrock}}, \bibinfo {author}
  {\bibfnamefont{K.~A.}\ \bibnamefont{Wendt}}, \bibinfo {author}
  {\bibfnamefont{S.}~\bibnamefont{Bacca}}, \bibinfo {author}
  {\bibfnamefont{N.}~\bibnamefont{Barnea}}, \bibinfo {author}
  {\bibfnamefont{B.}~\bibnamefont{Carlsson}}, \bibinfo {author}
  {\bibfnamefont{C.}~\bibnamefont{Drischler}}, \bibinfo {author}
  {\bibfnamefont{K.}~\bibnamefont{Hebeler}}, \bibinfo {author}
  {\bibfnamefont{M.}~\bibnamefont{Hjorth-Jensen}}, \bibinfo {author}
  {\bibfnamefont{M.}~\bibnamefont{Miorelli}}, \bibinfo {author}
  {\bibfnamefont{G.}~\bibnamefont{Orlandini}}, \bibinfo {author}
  {\bibfnamefont{A.}~\bibnamefont{Schwenk}},\ and\ \bibinfo {author}
  {\bibfnamefont{J.}~\bibnamefont{Simonis}},\ }%
  \bibfield{journal}{%
  \bibinfo {journal} {Nat. Phys.}\ }%
  \textbf{\bibinfo {volume} {12}},\ \bibinfo {pages} {186} (\bibinfo {year}
  {2016}),\ \url{http://dx.doi.org/10.1038/nphys3529}%
  \bibAnnoteFile{NoStop}{Hagen16}%
\bibitem{ReiNaz16}%
  \BibitemOpen
  \bibfield{author}{%
  \bibinfo {author} {\bibfnamefont{P.-G.}\ \bibnamefont{Reinhard}}\ and\
  \bibinfo {author} {\bibfnamefont{W.}~\bibnamefont{Nazarewicz}},\ }%
  \bibfield{journal}{%
  \Doi{10.1103/PhysRevC.93.051303}{\bibinfo {journal} {Phys. Rev. C}}\ }%
  \textbf{\bibinfo {volume} {93}},\ \bibinfo {pages} {051303} (\bibinfo {month}
  {May}\ \bibinfo {year} {2016}),\
  \url{https://link.aps.org/doi/10.1103/PhysRevC.93.051303}%
  \bibAnnoteFile{NoStop}{ReiNaz16}%
\bibitem{ang13}%
  \BibitemOpen
  \bibfield{author}{%
  \bibinfo {author} {\bibfnamefont{I.}~\bibnamefont{Angeli}}\ and\ \bibinfo
  {author} {\bibfnamefont{K.~P.}\ \bibnamefont{Marinova}},\ }%
  \bibfield{journal}{%
  \bibinfo {journal} {Atomic Data and Nuclear Data Tables}\ }%
  \textbf{\bibinfo {volume} {99}},\ \bibinfo {pages} {69} (\bibinfo {year}
  {2013})%
  \bibAnnoteFile{NoStop}{ang13}%
\bibitem{Sun17}%
  \BibitemOpen
  \bibfield{author}{%
  \bibinfo {author} {\bibfnamefont{B.~H.}\ \bibnamefont{Sun}}, \bibinfo
  {author} {\bibfnamefont{C.~Y.}\ \bibnamefont{Liu}},\ and\ \bibinfo {author}
  {\bibfnamefont{H.~X.}\ \bibnamefont{Wang}},\ }%
  \bibfield{journal}{%
  \Doi{10.1103/PhysRevC.95.014307}{\bibinfo {journal} {Phys. Rev. C}}\ }%
  \textbf{\bibinfo {volume} {95}},\ \bibinfo {pages} {014307} (\bibinfo {month}
  {Jan}\ \bibinfo {year} {2017}),\
  \url{https://link.aps.org/doi/10.1103/PhysRevC.95.014307}%
  \bibAnnoteFile{NoStop}{Sun17}%
\bibitem{DeWitte}%
  \BibitemOpen
  \bibfield{author}{%
  \bibinfo {author} {\bibfnamefont{H.}~\bibnamefont{DeWitte}} \emph{et~al.},\
  }%
  \bibfield{journal}{%
  \Doi{10.1103/PhysRevLett.98.112502}{\bibinfo {journal} {Phys. Rev. Lett.}}\
  }%
  \textbf{\bibinfo {volume} {98}},\ \bibinfo {pages} {112502} (\bibinfo {month}
  {Mar}\ \bibinfo {year} {2007}),\
  \url{http://link.aps.org/doi/10.1103/PhysRevLett.98.112502}%
  \bibAnnoteFile{NoStop}{DeWitte}%
\bibitem{GarciaRuiz16}%
  \BibitemOpen
  \bibfield{author}{%
  \bibinfo {author} {\bibfnamefont{G.}~\bibnamefont{Ruiz}} \emph{et~al.},\ }%
  \bibfield{journal}{%
  \bibinfo {journal} {Nat. Phys.}\ }%
  \textbf{\bibinfo {volume} {12}},\ \bibinfo {pages} {594} (\bibinfo {year}
  {2016}),\ \url{http://dx.doi.org/10.1038/nphys3645}%
  \bibAnnoteFile{NoStop}{GarciaRuiz16}%
\bibitem{Rossi15}%
  \BibitemOpen
  \bibfield{author}{%
  \bibinfo {author} {\bibfnamefont{D.~M.}\ \bibnamefont{Rossi}}, \bibinfo
  {author} {\bibfnamefont{K.}~\bibnamefont{Minamisono}}, \bibinfo {author}
  {\bibfnamefont{H.~B.}\ \bibnamefont{Asberry}}, \bibinfo {author}
  {\bibfnamefont{G.}~\bibnamefont{Bollen}}, \bibinfo {author}
  {\bibfnamefont{B.~A.}\ \bibnamefont{Brown}}, \bibinfo {author}
  {\bibfnamefont{K.}~\bibnamefont{Cooper}}, \bibinfo {author}
  {\bibfnamefont{B.}~\bibnamefont{Isherwood}}, \bibinfo {author}
  {\bibfnamefont{P.~F.}\ \bibnamefont{Mantica}}, \bibinfo {author}
  {\bibfnamefont{A.}~\bibnamefont{Miller}}, \bibinfo {author}
  {\bibfnamefont{D.~J.}\ \bibnamefont{Morrissey}}, \bibinfo {author}
  {\bibfnamefont{R.}~\bibnamefont{Ringle}}, \bibinfo {author}
  {\bibfnamefont{J.~A.}\ \bibnamefont{Rodriguez}}, \bibinfo {author}
  {\bibfnamefont{C.~A.}\ \bibnamefont{Ryder}}, \bibinfo {author}
  {\bibfnamefont{A.}~\bibnamefont{Smith}}, \bibinfo {author}
  {\bibfnamefont{R.}~\bibnamefont{Strum}},\ and\ \bibinfo {author}
  {\bibfnamefont{C.}~\bibnamefont{Sumithrarachchi}},\ }%
  \bibfield{journal}{%
  \Doi{10.1103/PhysRevC.92.014305}{\bibinfo {journal} {Phys. Rev. C}}\ }%
  \textbf{\bibinfo {volume} {92}},\ \bibinfo {pages} {014305} (\bibinfo {month}
  {Jul}\ \bibinfo {year} {2015}),\
  \url{http://link.aps.org/doi/10.1103/PhysRevC.92.014305}%
  \bibAnnoteFile{NoStop}{Rossi15}%
\bibitem{Minamisono16}%
  \BibitemOpen
  \bibfield{author}{%
  \bibinfo {author} {\bibfnamefont{K.}~\bibnamefont{Minamisono}}, \bibinfo
  {author} {\bibfnamefont{D.~M.}\ \bibnamefont{Rossi}}, \bibinfo {author}
  {\bibfnamefont{R.}~\bibnamefont{Beerwerth}}, \bibinfo {author}
  {\bibfnamefont{S.}~\bibnamefont{Fritzsche}}, \bibinfo {author}
  {\bibfnamefont{D.}~\bibnamefont{Garand}}, \bibinfo {author}
  {\bibfnamefont{A.}~\bibnamefont{Klose}}, \bibinfo {author}
  {\bibfnamefont{Y.}~\bibnamefont{Liu}}, \bibinfo {author}
  {\bibfnamefont{B.}~\bibnamefont{Maa\ss{}}}, \bibinfo {author}
  {\bibfnamefont{P.~F.}\ \bibnamefont{Mantica}}, \bibinfo {author}
  {\bibfnamefont{A.~J.}\ \bibnamefont{Miller}}, \bibinfo {author}
  {\bibfnamefont{P.}~\bibnamefont{M\"uller}}, \bibinfo {author}
  {\bibfnamefont{W.}~\bibnamefont{Nazarewicz}}, \bibinfo {author}
  {\bibfnamefont{W.}~\bibnamefont{N\"ortersh\"auser}}, \bibinfo {author}
  {\bibfnamefont{E.}~\bibnamefont{Olsen}}, \bibinfo {author}
  {\bibfnamefont{M.~R.}\ \bibnamefont{Pearson}}, \bibinfo {author}
  {\bibfnamefont{P.-G.}\ \bibnamefont{Reinhard}}, \bibinfo {author}
  {\bibfnamefont{E.~E.}\ \bibnamefont{Saperstein}}, \bibinfo {author}
  {\bibfnamefont{C.}~\bibnamefont{Sumithrarachchi}},\ and\ \bibinfo {author}
  {\bibfnamefont{S.~V.}\ \bibnamefont{Tolokonnikov}},\ }%
  \bibfield{journal}{%
  \Doi{10.1103/PhysRevLett.117.252501}{\bibinfo {journal} {Phys. Rev. Lett.}}\
  }%
  \textbf{\bibinfo {volume} {117}},\ \bibinfo {pages} {252501} (\bibinfo
  {month} {Dec}\ \bibinfo {year} {2016}),\
  \url{http://link.aps.org/doi/10.1103/PhysRevLett.117.252501}%
  \bibAnnoteFile{NoStop}{Minamisono16}%
\bibitem{(Ben03)}%
  \BibitemOpen
  \bibfield{author}{%
  \bibinfo {author} {\bibfnamefont{M.}~\bibnamefont{Bender}}, \bibinfo {author}
  {\bibfnamefont{P.-H.}\ \bibnamefont{Heenen}},\ and\ \bibinfo {author}
  {\bibfnamefont{P.-G.}\ \bibnamefont{Reinhard}},\ }%
  \bibfield{journal}{%
  \Doi{10.1103/RevModPhys.75.121}{\bibinfo {journal} {Rev. Mod. Phys.}}\ }%
  \textbf{\bibinfo {volume} {75}},\ \bibinfo {pages} {121} (\bibinfo {month}
  {Jan}\ \bibinfo {year} {2003}),\
  \url{http://link.aps.org/doi/10.1103/RevModPhys.75.121}%
  \bibAnnoteFile{NoStop}{(Ben03)}%
\bibitem{Patyk99}%
  \BibitemOpen
  \bibfield{author}{%
  \bibinfo {author} {\bibfnamefont{Z.}~\bibnamefont{Patyk}}, \bibinfo {author}
  {\bibfnamefont{A.}~\bibnamefont{Baran}}, \bibinfo {author}
  {\bibfnamefont{J.~F.}\ \bibnamefont{Berger}}, \bibinfo {author}
  {\bibfnamefont{J.}~\bibnamefont{Decharg\'e}}, \bibinfo {author}
  {\bibfnamefont{J.}~\bibnamefont{Dobaczewski}}, \bibinfo {author}
  {\bibfnamefont{P.}~\bibnamefont{Ring}},\ and\ \bibinfo {author}
  {\bibfnamefont{A.}~\bibnamefont{Sobiczewski}},\ }%
  \bibfield{journal}{%
  \Doi{10.1103/PhysRevC.59.704}{\bibinfo {journal} {Phys. Rev. C}}\ }%
  \textbf{\bibinfo {volume} {59}},\ \bibinfo {pages} {704} (\bibinfo {month}
  {Feb}\ \bibinfo {year} {1999}),\
  \url{https://link.aps.org/doi/10.1103/PhysRevC.59.704}%
  \bibAnnoteFile{NoStop}{Patyk99}%
\bibitem{Goriely09}%
  \BibitemOpen
  \bibfield{author}{%
  \bibinfo {author} {\bibfnamefont{S.}~\bibnamefont{Goriely}}, \bibinfo
  {author} {\bibfnamefont{S.}~\bibnamefont{Hilaire}}, \bibinfo {author}
  {\bibfnamefont{M.}~\bibnamefont{Girod}},\ and\ \bibinfo {author}
  {\bibfnamefont{S.}~\bibnamefont{P\'eru}},\ }%
  \bibfield{journal}{%
  \Doi{10.1103/PhysRevLett.102.242501}{\bibinfo {journal} {Phys. Rev. Lett.}}\
  }%
  \textbf{\bibinfo {volume} {102}},\ \bibinfo {pages} {242501} (\bibinfo
  {month} {Jun}\ \bibinfo {year} {2009}),\
  \url{https://link.aps.org/doi/10.1103/PhysRevLett.102.242501}%
  \bibAnnoteFile{NoStop}{Goriely09}%
\bibitem{Kortelainen_2010}%
  \BibitemOpen
  \bibfield{author}{%
  \bibinfo {author} {\bibfnamefont{M.}~\bibnamefont{Kortelainen}}, \bibinfo
  {author} {\bibfnamefont{T.}~\bibnamefont{Lesinski}}, \bibinfo {author}
  {\bibfnamefont{J.}~\bibnamefont{Mor\'e}}, \bibinfo {author}
  {\bibfnamefont{W.}~\bibnamefont{Nazarewicz}}, \bibinfo {author}
  {\bibfnamefont{J.}~\bibnamefont{Sarich}}, \bibinfo {author}
  {\bibfnamefont{N.}~\bibnamefont{Schunck}}, \bibinfo {author}
  {\bibfnamefont{M.~V.}\ \bibnamefont{Stoitsov}},\ and\ \bibinfo {author}
  {\bibfnamefont{S.}~\bibnamefont{Wild}},\ }%
  \bibfield{journal}{%
  \Doi{10.1103/PhysRevC.82.024313}{\bibinfo {journal} {Phys. Rev. C}}\ }%
  \textbf{\bibinfo {volume} {82}},\ \bibinfo {pages} {024313} (\bibinfo {month}
  {Aug}\ \bibinfo {year} {2010}),\
  \url{http://link.aps.org/doi/10.1103/PhysRevC.82.024313}%
  \bibAnnoteFile{NoStop}{Kortelainen_2010}%
\bibitem{Utama16}%
  \BibitemOpen
  \bibfield{author}{%
  \bibinfo {author} {\bibfnamefont{R.}~\bibnamefont{Utama}}, \bibinfo {author}
  {\bibfnamefont{W.-C.}\ \bibnamefont{Chen}},\ and\ \bibinfo {author}
  {\bibfnamefont{J.}~\bibnamefont{Piekarewicz}},\ }%
  \bibfield{journal}{%
  \bibinfo {journal} {J. Phys. G}\ }%
  \textbf{\bibinfo {volume} {43}},\ \bibinfo {pages} {114002} (\bibinfo {year}
  {2016}),\ \url{http://stacks.iop.org/0954-3899/43/i=11/a=114002}%
  \bibAnnoteFile{NoStop}{Utama16}%
\bibitem{Smirnov88}%
  \BibitemOpen
  \bibfield{author}{%
  \bibinfo {author} {\bibfnamefont{A.~V.}\ \bibnamefont{Smirnov}}, \bibinfo
  {author} {\bibfnamefont{S.~V.}\ \bibnamefont{Tolokonnikov}},\ and\ \bibinfo
  {author} {\bibfnamefont{S.~A.}\ \bibnamefont{Fayans}},\ }%
  \bibfield{journal}{%
  \bibinfo {journal} {Sov. J. Nucl. Phys.}\ }%
  \textbf{\bibinfo {volume} {48}},\ \bibinfo {pages} {995} (\bibinfo {year}
  {1988})%
  \bibAnnoteFile{NoStop}{Smirnov88}%
\bibitem{Fay94}%
  \BibitemOpen
  \bibfield{author}{%
  \bibinfo {author} {\bibfnamefont{S.}~\bibnamefont{Fayans}}, \bibinfo {author}
  {\bibfnamefont{E.}~\bibnamefont{Trykov}},\ and\ \bibinfo {author}
  {\bibfnamefont{D.}~\bibnamefont{Zawischa}},\ }%
  \bibfield{journal}{%
  \Doi{http://dx.doi.org/10.1016/0375-9474(94)90392-1}{\bibinfo {journal}
  {Nucl. Phys. A}}\ }%
  \textbf{\bibinfo {volume} {568}},\ \bibinfo {pages} {523} (\bibinfo {year}
  {1994}),\
  \url{http://www.sciencedirect.com/science/article/pii/0375947494903921}%
  \bibAnnoteFile{NoStop}{Fay94}%
\bibitem{Borzov1996}%
  \BibitemOpen
  \bibfield{author}{%
  \bibinfo {author} {\bibfnamefont{I.~N.}\ \bibnamefont{Borzov}}, \bibinfo
  {author} {\bibfnamefont{S.~A.}\ \bibnamefont{Fayans}}, \bibinfo {author}
  {\bibfnamefont{E.}~\bibnamefont{Kr{\"o}mer}},\ and\ \bibinfo {author}
  {\bibfnamefont{D.}~\bibnamefont{Zawischa}},\ }%
  \bibfield{journal}{%
  \Doi{10.1007/BF02769674}{\bibinfo {journal} {Z. Phys. A}}\ }%
  \textbf{\bibinfo {volume} {355}},\ \bibinfo {pages} {117} (\bibinfo {year}
  {1996}),\ ISSN \bibinfo {issn} {0939-7922},\
  \url{http://dx.doi.org/10.1007/BF02769674}%
  \bibAnnoteFile{NoStop}{Borzov1996}%
\bibitem{Fayans1998}%
  \BibitemOpen
  \bibfield{author}{%
  \bibinfo {author} {\bibfnamefont{S.~A.}\ \bibnamefont{Fayans}},\ }%
  \bibfield{journal}{%
  \Doi{10.1134/1.567841}{\bibinfo {journal} {JETP Lett.}}\ }%
  \textbf{\bibinfo {volume} {68}},\ \bibinfo {pages} {169} (\bibinfo {year}
  {1998}),\ \url{http://dx.doi.org/10.1134/1.567841}%
  \bibAnnoteFile{NoStop}{Fayans1998}%
\bibitem{Fayans2000}%
  \BibitemOpen
  \bibfield{author}{%
  \bibinfo {author} {\bibfnamefont{S.}~\bibnamefont{Fayans}}, \bibinfo {author}
  {\bibfnamefont{S.}~\bibnamefont{Tolokonnikov}}, \bibinfo {author}
  {\bibfnamefont{E.}~\bibnamefont{Trykov}},\ and\ \bibinfo {author}
  {\bibfnamefont{D.}~\bibnamefont{Zawischa}},\ }%
  \bibfield{journal}{%
  \bibinfo {journal} {Nucl. Phys. A}\ }%
  \textbf{\bibinfo {volume} {676}},\ \bibinfo {pages} {49} (\bibinfo {year}
  {2000})%
  \bibAnnoteFile{NoStop}{Fayans2000}%
\bibitem{Fay1994}%
  \BibitemOpen
  \bibfield{author}{%
  \bibinfo {author} {\bibfnamefont{S.}~\bibnamefont{Fayans}}, \bibinfo {author}
  {\bibfnamefont{S.}~\bibnamefont{Tolokonnikov}}, \bibinfo {author}
  {\bibfnamefont{E.}~\bibnamefont{Trykov}},\ and\ \bibinfo {author}
  {\bibfnamefont{D.}~\bibnamefont{Zawischa}},\ }%
  \bibfield{journal}{%
  \Doi{http://dx.doi.org/10.1016/0370-2693(94)91334-X}{\bibinfo {journal}
  {Phys. Lett. B}}\ }%
  \textbf{\bibinfo {volume} {338}},\ \bibinfo {pages} {1 } (\bibinfo {year}
  {1994}),\ ISSN \bibinfo {issn} {0370-2693},\
  \url{http://www.sciencedirect.com/science/article/pii/037026939491334X}%
  \bibAnnoteFile{NoStop}{Fay1994}%
\bibitem{FayZ96}%
  \BibitemOpen
  \bibfield{author}{%
  \bibinfo {author} {\bibfnamefont{S.}~\bibnamefont{Fayans}}\ and\ \bibinfo
  {author} {\bibfnamefont{D.}~\bibnamefont{Zawischa}},\ }%
  \bibfield{journal}{%
  \Doi{http://dx.doi.org/10.1016/0370-2693(96)00716-2}{\bibinfo {journal}
  {Phys. Lett. B}}\ }%
  \textbf{\bibinfo {volume} {383}},\ \bibinfo {pages} {19} (\bibinfo {year}
  {1996}),\ ISSN \bibinfo {issn} {0370-2693},\
  \url{http://www.sciencedirect.com/science/article/pii/0370269396007162}%
  \bibAnnoteFile{NoStop}{FayZ96}%
\bibitem{Engel_1975}%
  \BibitemOpen
  \bibfield{author}{%
  \bibinfo {author} {\bibfnamefont{Y.~M.}\ \bibnamefont{Engel}}, \bibinfo
  {author} {\bibfnamefont{D.~M.}\ \bibnamefont{Brink}}, \bibinfo {author}
  {\bibfnamefont{K.}~\bibnamefont{Goeke}}, \bibinfo {author}
  {\bibfnamefont{S.~J.}\ \bibnamefont{Krieger}},\ and\ \bibinfo {author}
  {\bibfnamefont{D.}~\bibnamefont{Vautherin}},\ }%
  \bibfield{journal}{%
  \bibinfo {journal} {Nucl. Phys. A}\ }%
  \textbf{\bibinfo {volume} {249}},\ \bibinfo {pages} {215} (\bibinfo {year}
  {1975})%
  \bibAnnoteFile{NoStop}{Engel_1975}%
\bibitem{Per04}%
  \BibitemOpen
  \bibfield{author}{%
  \bibinfo {author} {\bibfnamefont{E.}~\bibnamefont{Perli{\'n}ska}}, \bibinfo
  {author} {\bibfnamefont{S.~G.}\ \bibnamefont{Rohozi{\'n}ski}}, \bibinfo
  {author} {\bibfnamefont{J.}~\bibnamefont{Dobaczewski}},\ and\ \bibinfo
  {author} {\bibfnamefont{W.}~\bibnamefont{Nazarewicz}},\ }%
  \bibfield{journal}{%
  \Doi{10.1103/PhysRevC.69.014316}{\bibinfo {journal} {Phys. Rev. C}}\ }%
  \textbf{\bibinfo {volume} {69}},\ \bibinfo {pages} {014316} (\bibinfo {month}
  {Jan}\ \bibinfo {year} {2004}),\
  \url{http://link.aps.org/doi/10.1103/PhysRevC.69.014316}%
  \bibAnnoteFile{NoStop}{Per04}%
\bibitem{Kluepfel_2009}%
  \BibitemOpen
  \bibfield{author}{%
  \bibinfo {author} {\bibfnamefont{P.}~\bibnamefont{Kl{\"{u}}pfel}}, \bibinfo
  {author} {\bibfnamefont{P.-G.}\ \bibnamefont{Reinhard}}, \bibinfo {author}
  {\bibfnamefont{T.~J.}\ \bibnamefont{B{\"{u}}rvenich}},\ and\ \bibinfo
  {author} {\bibfnamefont{J.~A.}\ \bibnamefont{Maruhn}},\ }%
  \bibfield{journal}{%
  \bibinfo {journal} {Phys. Rev. C}\ }%
  \textbf{\bibinfo {volume} {79}},\ \bibinfo {pages} {034310} (\bibinfo {month}
  {Mar}\ \bibinfo {year} {2009})%
  \bibAnnoteFile{NoStop}{Kluepfel_2009}%
\bibitem{Dob01}%
  \BibitemOpen
  \bibfield{author}{%
  \bibinfo {author} {\bibfnamefont{J.}~\bibnamefont{Dobaczewski}}, \bibinfo
  {author} {\bibfnamefont{W.}~\bibnamefont{Nazarewicz}},\ and\ \bibinfo
  {author} {\bibfnamefont{P.-G.}\ \bibnamefont{Reinhard}},\ }%
  \bibfield{journal}{%
  \Doi{http://dx.doi.org/10.1016/S0375-9474(01)00993-9}{\bibinfo {journal}
  {Nucl. Phys. A}}\ }%
  \textbf{\bibinfo {volume} {693}},\ \bibinfo {pages} {361} (\bibinfo {year}
  {2001}),\ ISSN \bibinfo {issn} {0375-9474},\
  \url{http://www.sciencedirect.com/science/article/pii/S0375947401009939}%
  \bibAnnoteFile{NoStop}{Dob01}%
\bibitem{Dob02a}%
  \BibitemOpen
  \bibfield{author}{%
  \bibinfo {author} {\bibfnamefont{J.}~\bibnamefont{Dobaczewski}}, \bibinfo
  {author} {\bibfnamefont{W.}~\bibnamefont{Nazarewicz}},\ and\ \bibinfo
  {author} {\bibfnamefont{M.~V.}\ \bibnamefont{Stoitsov}},\ }%
  \bibfield{journal}{%
  \bibinfo {journal} {Eur. Phys. J. A}\ }%
  \textbf{\bibinfo {volume} {15}},\ \bibinfo {pages} {21} (\bibinfo {year}
  {2002})%
  \bibAnnoteFile{NoStop}{Dob02a}%
\bibitem{Bertsch09}%
  \BibitemOpen
  \bibfield{author}{%
  \bibinfo {author} {\bibfnamefont{G.~F.}\ \bibnamefont{Bertsch}}, \bibinfo
  {author} {\bibfnamefont{C.~A.}\ \bibnamefont{Bertulani}}, \bibinfo {author}
  {\bibfnamefont{W.}~\bibnamefont{Nazarewicz}}, \bibinfo {author}
  {\bibfnamefont{N.}~\bibnamefont{Schunck}},\ and\ \bibinfo {author}
  {\bibfnamefont{M.~V.}\ \bibnamefont{Stoitsov}},\ }%
  \bibfield{journal}{%
  \Doi{10.1103/PhysRevC.79.034306}{\bibinfo {journal} {Phys. Rev. C}}\ }%
  \textbf{\bibinfo {volume} {79}},\ \bibinfo {pages} {034306} (\bibinfo {month}
  {Mar}\ \bibinfo {year} {2009}),\
  \url{http://link.aps.org/doi/10.1103/PhysRevC.79.034306}%
  \bibAnnoteFile{NoStop}{Bertsch09}%
\bibitem{Tolok15}%
  \BibitemOpen
  \bibfield{author}{%
  \bibinfo {author} {\bibfnamefont{S.~V.}\ \bibnamefont{Tolokonnikov}},
  \bibinfo {author} {\bibfnamefont{I.~N.}\ \bibnamefont{Borzov}}, \bibinfo
  {author} {\bibfnamefont{M.}~\bibnamefont{Kortelainen}}, \bibinfo {author}
  {\bibfnamefont{Y.~S.}\ \bibnamefont{Lutostansky}},\ and\ \bibinfo {author}
  {\bibfnamefont{E.~E.}\ \bibnamefont{Saperstein}},\ }%
  \bibfield{journal}{%
  \bibinfo {journal} {J. Phys. G}\ }%
  \textbf{\bibinfo {volume} {42}},\ \bibinfo {pages} {075102} (\bibinfo {year}
  {2015})%
  \bibAnnoteFile{NoStop}{Tolok15}%
\bibitem{Kromer95}%
  \BibitemOpen
  \bibfield{author}{%
  \bibinfo {author} {\bibfnamefont{E.}~\bibnamefont{Kr{\"o}mer}}, \bibinfo
  {author} {\bibfnamefont{S.}~\bibnamefont{Tolokonnikov}}, \bibinfo {author}
  {\bibfnamefont{S.}~\bibnamefont{Fayans}},\ and\ \bibinfo {author}
  {\bibfnamefont{D.}~\bibnamefont{Zawischa}},\ }%
  \bibfield{journal}{%
  \Doi{http://dx.doi.org/10.1016/0370-2693(95)01216-D}{\bibinfo {journal}
  {Phys. Lett. B}}\ }%
  \textbf{\bibinfo {volume} {363}},\ \bibinfo {pages} {12} (\bibinfo {year}
  {1995}),\ ISSN \bibinfo {issn} {0370-2693},\
  \url{http://www.sciencedirect.com/science/article/pii/037026939501216D}%
  \bibAnnoteFile{NoStop}{Kromer95}%
\bibitem{Horen96}%
  \BibitemOpen
  \bibfield{author}{%
  \bibinfo {author} {\bibfnamefont{D.}~\bibnamefont{Horen}}, \bibinfo {author}
  {\bibfnamefont{G.}~\bibnamefont{Satchler}}, \bibinfo {author}
  {\bibfnamefont{S.}~\bibnamefont{Fayans}},\ and\ \bibinfo {author}
  {\bibfnamefont{E.}~\bibnamefont{Trykov}},\ }%
  \bibfield{journal}{%
  \Doi{http://dx.doi.org/10.1016/0375-9474(96)00032-2}{\bibinfo {journal}
  {Nucl. Phys. A}}\ }%
  \textbf{\bibinfo {volume} {600}},\ \bibinfo {pages} {193} (\bibinfo {year}
  {1996}),\
  \url{http://www.sciencedirect.com/science/article/pii/0375947496000322}%
  \bibAnnoteFile{NoStop}{Horen96}%
\bibitem{Tolokonnikov2010}%
  \BibitemOpen
  \bibfield{author}{%
  \bibinfo {author} {\bibfnamefont{S.~V.}\ \bibnamefont{Tolokonnikov}}\ and\
  \bibinfo {author} {\bibfnamefont{E.~E.}\ \bibnamefont{Saperstein}},\ }%
  \bibfield{journal}{%
  \bibinfo {journal} {Phys. At. Nucl.}\ }%
  \textbf{\bibinfo {volume} {73}},\ \bibinfo {pages} {1684} (\bibinfo {year}
  {2010})%
  \bibAnnoteFile{NoStop}{Tolokonnikov2010}%
\bibitem{Erler10}%
  \BibitemOpen
  \bibfield{author}{%
  \bibinfo {author} {\bibfnamefont{J.}~\bibnamefont{Erler}}, \bibinfo {author}
  {\bibfnamefont{P.}~\bibnamefont{Kl\"upfel}},\ and\ \bibinfo {author}
  {\bibfnamefont{P.-G.}\ \bibnamefont{Reinhard}},\ }%
  \bibfield{journal}{%
  \bibinfo {journal} {Phys. Rev. C}\ }%
  \textbf{\bibinfo {volume} {82}},\ \bibinfo {pages} {044307} (\bibinfo {year}
  {2010}),\ \url{http://link.aps.org/doi/10.1103/PhysRevC.82.044307}%
  \bibAnnoteFile{NoStop}{Erler10}%
\bibitem{Ben07a}%
  \BibitemOpen
  \bibfield{author}{%
  \bibinfo {author} {\bibfnamefont{T.}~\bibnamefont{Lesinski}}, \bibinfo
  {author} {\bibfnamefont{M.}~\bibnamefont{Bender}}, \bibinfo {author}
  {\bibfnamefont{K.}~\bibnamefont{Bennaceur}}, \bibinfo {author}
  {\bibfnamefont{T.}~\bibnamefont{Duguet}},\ and\ \bibinfo {author}
  {\bibfnamefont{J.}~\bibnamefont{Meyer}},\ }%
  \bibfield{journal}{%
  \bibinfo {journal} {Phys. Rev. C}\ }%
  \textbf{\bibinfo {volume} {76}},\ \bibinfo {pages} {014312} (\bibinfo {year}
  {2007})%
  \bibAnnoteFile{NoStop}{Ben07a}%
\bibitem{Kor14a}%
  \BibitemOpen
  \bibfield{author}{%
  \bibinfo {author} {\bibfnamefont{M.}~\bibnamefont{Kortelainen}}, \bibinfo
  {author} {\bibfnamefont{J.}~\bibnamefont{McDonnell}}, \bibinfo {author}
  {\bibfnamefont{W.}~\bibnamefont{Nazarewicz}}, \bibinfo {author}
  {\bibfnamefont{E.}~\bibnamefont{Olsen}}, \bibinfo {author}
  {\bibfnamefont{P.-G.}\ \bibnamefont{Reinhard}}, \bibinfo {author}
  {\bibfnamefont{J.}~\bibnamefont{Sarich}}, \bibinfo {author}
  {\bibfnamefont{N.}~\bibnamefont{Schunck}}, \bibinfo {author}
  {\bibfnamefont{S.~M.}\ \bibnamefont{Wild}}, \bibinfo {author}
  {\bibfnamefont{D.}~\bibnamefont{Davesne}}, \bibinfo {author}
  {\bibfnamefont{J.}~\bibnamefont{Erler}},\ and\ \bibinfo {author}
  {\bibfnamefont{A.}~\bibnamefont{Pastore}},\ }%
  \bibfield{journal}{%
  \bibinfo {journal} {Phys. Rev. C}\ }%
  \textbf{\bibinfo {volume} {89}},\ \bibinfo {pages} {054314} (\bibinfo {year}
  {2014})%
  \bibAnnoteFile{NoStop}{Kor14a}%
\bibitem{Kim92}%
  \BibitemOpen
  \bibfield{author}{%
  \bibinfo {author} {\bibfnamefont{W.}~\bibnamefont{Kim}}, \bibinfo {author}
  {\bibfnamefont{J.~P.}\ \bibnamefont{Connelly}}, \bibinfo {author}
  {\bibfnamefont{J.~H.}\ \bibnamefont{Heisenberg}}, \bibinfo {author}
  {\bibfnamefont{F.~W.}\ \bibnamefont{Hersman}}, \bibinfo {author}
  {\bibfnamefont{T.~E.}\ \bibnamefont{Milliman}}, \bibinfo {author}
  {\bibfnamefont{J.~E.}\ \bibnamefont{Wise}}, \bibinfo {author}
  {\bibfnamefont{C.~N.}\ \bibnamefont{Papanicolas}}, \bibinfo {author}
  {\bibfnamefont{S.~A.}\ \bibnamefont{Fayans}},\ and\ \bibinfo {author}
  {\bibfnamefont{A.~P.}\ \bibnamefont{Platonov}},\ }%
  \bibfield{journal}{%
  \Doi{10.1103/PhysRevC.46.1656}{\bibinfo {journal} {Phys. Rev. C}}\ }%
  \textbf{\bibinfo {volume} {46}},\ \bibinfo {pages} {1656} (\bibinfo {month}
  {Nov}\ \bibinfo {year} {1992}),\
  \url{http://link.aps.org/doi/10.1103/PhysRevC.46.1656}%
  \bibAnnoteFile{NoStop}{Kim92}%
\bibitem{Gnezdilov14}%
  \BibitemOpen
  \bibfield{author}{%
  \bibinfo {author} {\bibfnamefont{N.~V.}\ \bibnamefont{Gnezdilov}}, \bibinfo
  {author} {\bibfnamefont{I.~N.}\ \bibnamefont{Borzov}}, \bibinfo {author}
  {\bibfnamefont{E.~E.}\ \bibnamefont{Saperstein}},\ and\ \bibinfo {author}
  {\bibfnamefont{S.~V.}\ \bibnamefont{Tolokonnikov}},\ }%
  \bibfield{journal}{%
  \Doi{10.1103/PhysRevC.89.034304}{\bibinfo {journal} {Phys. Rev. C}}\ }%
  \textbf{\bibinfo {volume} {89}},\ \bibinfo {pages} {034304} (\bibinfo {year}
  {2014}),\ \url{http://link.aps.org/doi/10.1103/PhysRevC.89.034304}%
  \bibAnnoteFile{NoStop}{Gnezdilov14}%
\bibitem{Saperstein2016}%
  \BibitemOpen
  \bibfield{author}{%
  \bibinfo {author} {\bibfnamefont{E.~E.}\ \bibnamefont{Saperstein}}, \bibinfo
  {author} {\bibfnamefont{I.~N.}\ \bibnamefont{Borzov}},\ and\ \bibinfo
  {author} {\bibfnamefont{S.~V.}\ \bibnamefont{Tolokonnikov}},\ }%
  \bibfield{journal}{%
  \Doi{10.1134/S0021364016160128}{\bibinfo {journal} {JETP Letters}}\ }%
  \textbf{\bibinfo {volume} {104}},\ \bibinfo {pages} {218} (\bibinfo {year}
  {2016}),\ ISSN \bibinfo {issn} {1090-6487},\
  \url{http://dx.doi.org/10.1134/S0021364016160128}%
  \bibAnnoteFile{NoStop}{Saperstein2016}%
\bibitem{Fri82a}%
  \BibitemOpen
  \bibfield{author}{%
  \bibinfo {author} {\bibfnamefont{J.}~\bibnamefont{Friedrich}}\ and\ \bibinfo
  {author} {\bibfnamefont{N.}~\bibnamefont{V{\"o}gler}},\ }%
  \bibfield{journal}{%
  \bibinfo {journal} {Nucl. Phys. A}\ }%
  \textbf{\bibinfo {volume} {373}},\ \bibinfo {pages} {192} (\bibinfo {year}
  {1982})%
  \bibAnnoteFile{NoStop}{Fri82a}%
\bibitem{Hinohara16}%
  \BibitemOpen
  \bibfield{author}{%
  \bibinfo {author} {\bibfnamefont{N.}~\bibnamefont{Hinohara}}\ and\ \bibinfo
  {author} {\bibfnamefont{W.}~\bibnamefont{Nazarewicz}},\ }%
  \bibfield{journal}{%
  \Doi{10.1103/PhysRevLett.116.152502}{\bibinfo {journal} {Phys. Rev. Lett.}}\
  }%
  \textbf{\bibinfo {volume} {116}},\ \bibinfo {pages} {152502} (\bibinfo
  {month} {Apr}\ \bibinfo {year} {2016}),\
  \url{http://link.aps.org/doi/10.1103/PhysRevLett.116.152502}%
  \bibAnnoteFile{NoStop}{Hinohara16}%
\bibitem{Nazarewicz2014}%
  \BibitemOpen
  \bibfield{author}{%
  \bibinfo {author} {\bibfnamefont{W.}~\bibnamefont{Nazarewicz}}, \bibinfo
  {author} {\bibfnamefont{P.-G.}\ \bibnamefont{Reinhard}}, \bibinfo {author}
  {\bibfnamefont{W.}~\bibnamefont{Satu{\l}a}},\ and\ \bibinfo {author}
  {\bibfnamefont{D.}~\bibnamefont{Vretenar}},\ }%
  \bibfield{journal}{%
  \bibinfo {journal} {Eur. Phys. J. A}\ }%
  \textbf{\bibinfo {volume} {50}},\ \bibinfo {pages} {20} (\bibinfo {year}
  {2014})%
  \bibAnnoteFile{NoStop}{Nazarewicz2014}%
\bibitem{Bev69aB}%
  \BibitemOpen
  \bibfield{author}{%
  \bibinfo {author} {\bibfnamefont{P.~R.}\ \bibnamefont{Bevington}}\ and\
  \bibinfo {author} {\bibfnamefont{D.~K.}\ \bibnamefont{Robinson}},\ }%
  \emph{\bibinfo {title} {Data Reduction and Error Analysis for the Physical
  Sciences}}\ (\bibinfo {publisher} {McGraw-Hill},\ \bibinfo {year} {2003})%
  \bibAnnoteFile{NoStop}{Bev69aB}%
\bibitem{Bra97aB}%
  \BibitemOpen
  \bibfield{author}{%
  \bibinfo {author} {\bibfnamefont{S.}~\bibnamefont{Brandt}},\ }%
  \emph{\bibinfo {title} {Statistical and computational methods in data
  analysis}}\ (\bibinfo {publisher} {Springer-Verlag},\ \bibinfo {address} {New
  York},\ \bibinfo {year} {1997})%
  \bibAnnoteFile{NoStop}{Bra97aB}%
\bibitem{Fri86a}%
  \BibitemOpen
  \bibfield{author}{%
  \bibinfo {author} {\bibfnamefont{J.}~\bibnamefont{Friedrich}}\ and\ \bibinfo
  {author} {\bibfnamefont{P.-G.}\ \bibnamefont{Reinhard}},\ }%
  \bibfield{journal}{%
  \bibinfo {journal} {Phys. Rev. C}\ }%
  \textbf{\bibinfo {volume} {33}},\ \bibinfo {pages} {335} (\bibinfo {year}
  {1986})%
  \bibAnnoteFile{NoStop}{Fri86a}%
\bibitem{Sam02a}%
  \BibitemOpen
  \bibfield{author}{%
  \bibinfo {author} {\bibfnamefont{M.}~\bibnamefont{Samyn}}, \bibinfo {author}
  {\bibfnamefont{S.}~\bibnamefont{Goriely}}, \bibinfo {author}
  {\bibfnamefont{P.-H.}\ \bibnamefont{Heenen}}, \bibinfo {author}
  {\bibfnamefont{J.~M.}\ \bibnamefont{Pearson}},\ and\ \bibinfo {author}
  {\bibfnamefont{F.}~\bibnamefont{Tondeur}},\ }%
  \bibfield{journal}{%
  \bibinfo {journal} {Nucl. Phys. A}\ }%
  \textbf{\bibinfo {volume} {700}},\ \bibinfo {pages} {142} (\bibinfo {year}
  {2002})%
  \bibAnnoteFile{NoStop}{Sam02a}%
\bibitem{Dob14a}%
  \BibitemOpen
  \bibfield{author}{%
  \bibinfo {author} {\bibfnamefont{J.}~\bibnamefont{Dobaczewski}}, \bibinfo
  {author} {\bibfnamefont{W.}~\bibnamefont{Nazarewicz}},\ and\ \bibinfo
  {author} {\bibfnamefont{P.-G.}\ \bibnamefont{Reinhard}},\ }%
  \bibfield{journal}{%
  \bibinfo {journal} {J. Phys. G}\ }%
  \textbf{\bibinfo {volume} {41}},\ \bibinfo {pages} {074001} (\bibinfo {year}
  {2014})%
  \bibAnnoteFile{NoStop}{Dob14a}%
\bibitem{Kluepfel_2008}%
  \BibitemOpen
  \bibfield{author}{%
  \bibinfo {author} {\bibfnamefont{P.}~\bibnamefont{Kl{\"{u}}pfel}}, \bibinfo
  {author} {\bibfnamefont{J.}~\bibnamefont{Erler}}, \bibinfo {author}
  {\bibfnamefont{P.-G.}\ \bibnamefont{Reinhard}},\ and\ \bibinfo {author}
  {\bibfnamefont{J.~A.}\ \bibnamefont{Maruhn}},\ }%
  \bibfield{journal}{%
  \bibinfo {journal} {Eur. Phys. J. A}\ }%
  \textbf{\bibinfo {volume} {37}},\ \bibinfo {pages} {343} (\bibinfo {year}
  {2008})%
  \bibAnnoteFile{NoStop}{Kluepfel_2008}%
\bibitem{NUBASE17}%
  \BibitemOpen
  \bibfield{author}{%
  \bibinfo {author} {\bibfnamefont{G.}~\bibnamefont{Audi}}, \bibinfo {author}
  {\bibfnamefont{F.}~\bibnamefont{Kondev}}, \bibinfo {author}
  {\bibfnamefont{M.}~\bibnamefont{Wang}}, \bibinfo {author}
  {\bibfnamefont{W.}~\bibnamefont{Huang}},\ and\ \bibinfo {author}
  {\bibfnamefont{S.}~\bibnamefont{Naimi}},\ }%
  \bibfield{journal}{%
  \bibinfo {journal} {Chin. Phys. C}\ }%
  \textbf{\bibinfo {volume} {41}},\ \bibinfo {pages} {030001} (\bibinfo {year}
  {2017}),\ \url{http://stacks.iop.org/1674-1137/41/i=3/a=030001}%
  \bibAnnoteFile{NoStop}{NUBASE17}%
\bibitem{Naz10a}%
  \BibitemOpen
  \bibfield{author}{%
  \bibinfo {author} {\bibfnamefont{P.-G.}\ \bibnamefont{Reinhard}}\ and\
  \bibinfo {author} {\bibfnamefont{W.}~\bibnamefont{Nazarewicz}},\ }%
  \bibfield{journal}{%
  \Doi{10.1103/PhysRevC.81.051303}{\bibinfo {journal} {Phys. Rev. C}}\ }%
  \textbf{\bibinfo {volume} {81}},\ \bibinfo {pages} {051303(R)} (\bibinfo
  {month} {May}\ \bibinfo {year} {2010}),\
  \url{http://link.aps.org/doi/10.1103/PhysRevC.81.051303}%
  \bibAnnoteFile{NoStop}{Naz10a}%
\bibitem{Erl14b}%
  \BibitemOpen
  \bibfield{author}{%
  \bibinfo {author} {\bibfnamefont{J.}~\bibnamefont{Erler}}\ and\ \bibinfo
  {author} {\bibfnamefont{P.-G.}\ \bibnamefont{Reinhard}},\ }%
  \bibfield{journal}{%
  \bibinfo {journal} {J. Phys. G}\ }%
  \textbf{\bibinfo {volume} {42}},\ \bibinfo {pages} {034026} (\bibinfo {year}
  {2015})%
  \bibAnnoteFile{NoStop}{Erl14b}%
\bibitem{Rei91aR}%
  \BibitemOpen
  \bibfield{author}{%
  \bibinfo {author} {\bibfnamefont{P.-G.}\ \bibnamefont{Reinhard}},\ }%
  in\ \emph{\bibinfo {booktitle} {Computational Nuclear Physics {I} - Nuclear
  Structure}},\ \bibinfo {editor} {edited by\ \bibinfo {editor}
  {\bibfnamefont{K.}~\bibnamefont{Langanke}}, \bibinfo {editor}
  {\bibfnamefont{S.}~\bibnamefont{Koonin}},\ and\ \bibinfo {editor}
  {\bibfnamefont{J.}~\bibnamefont{Maruhn}}}\ (\bibinfo {publisher} {Springer},\
  \bibinfo {address} {Berlin},\ \bibinfo {year} {1991})\ p.~\bibinfo {pages}
  {28}%
  \bibAnnoteFile{NoStop}{Rei91aR}%
\bibitem{Rei82a}%
  \BibitemOpen
  \bibfield{author}{%
  \bibinfo {author} {\bibfnamefont{P.-G.}\ \bibnamefont{Reinhard}}\ and\
  \bibinfo {author} {\bibfnamefont{R.}~\bibnamefont{Cusson}},\ }%
  \bibfield{journal}{%
  \bibinfo {journal} {Nucl. Phys. A}\ }%
  \textbf{\bibinfo {volume} {378}},\ \bibinfo {pages} {418} (\bibinfo {year}
  {1982})%
  \bibAnnoteFile{NoStop}{Rei82a}%
\bibitem{Kri90a}%
  \BibitemOpen
  \bibfield{author}{%
  \bibinfo {author} {\bibfnamefont{S.~J.}\ \bibnamefont{Krieger}}, \bibinfo
  {author} {\bibfnamefont{P.}~\bibnamefont{Bonche}}, \bibinfo {author}
  {\bibfnamefont{H.}~\bibnamefont{Flocard}}, \bibinfo {author}
  {\bibfnamefont{P.}~\bibnamefont{Quentin}},\ and\ \bibinfo {author}
  {\bibfnamefont{M.~S.}\ \bibnamefont{Weiss}},\ }%
  \bibfield{journal}{%
  \bibinfo {journal} {Nucl. Phys. A}\ }%
  \textbf{\bibinfo {volume} {517}},\ \bibinfo {pages} {275} (\bibinfo {year}
  {1990})%
  \bibAnnoteFile{NoStop}{Kri90a}%
\bibitem{Erl08a}%
  \BibitemOpen
  \bibfield{author}{%
  \bibinfo {author} {\bibfnamefont{J.}~\bibnamefont{Erler}}, \bibinfo {author}
  {\bibfnamefont{P.}~\bibnamefont{Kl\"upfel}},\ and\ \bibinfo {author}
  {\bibfnamefont{P.-G.}\ \bibnamefont{Reinhard}},\ }%
  \bibfield{journal}{%
  \bibinfo {journal} {Eur. Phys. J. A}\ }%
  \textbf{\bibinfo {volume} {37}},\ \bibinfo {pages} {81} (\bibinfo {year}
  {2008}),\ \bibinfo {note} {http://www.arxiv.org/abs/0811.1442},\
  \url{http://dx.doi.org/10.1140/epja/i2008-10615-5}%
  \bibAnnoteFile{NoStop}{Erl08a}%
\bibitem{Dug01}%
  \BibitemOpen
  \bibfield{author}{%
  \bibinfo {author} {\bibfnamefont{T.}~\bibnamefont{Duguet}}, \bibinfo {author}
  {\bibfnamefont{P.}~\bibnamefont{Bonche}}, \bibinfo {author}
  {\bibfnamefont{P.-H.}\ \bibnamefont{Heenen}},\ and\ \bibinfo {author}
  {\bibfnamefont{J.}~\bibnamefont{Meyer}},\ }%
  \bibfield{journal}{%
  \Doi{10.1103/PhysRevC.65.014310}{\bibinfo {journal} {Phys. Rev. C}}\ }%
  \textbf{\bibinfo {volume} {65}},\ \bibinfo {pages} {014310} (\bibinfo {month}
  {Dec}\ \bibinfo {year} {2001}),\
  \url{http://link.aps.org/doi/10.1103/PhysRevC.65.014310}%
  \bibAnnoteFile{NoStop}{Dug01}%
\bibitem{Sch10}%
  \BibitemOpen
  \bibfield{author}{%
  \bibinfo {author} {\bibfnamefont{N.}~\bibnamefont{Schunck}}, \bibinfo
  {author} {\bibfnamefont{J.}~\bibnamefont{Dobaczewski}}, \bibinfo {author}
  {\bibfnamefont{J.}~\bibnamefont{McDonnell}}, \bibinfo {author}
  {\bibfnamefont{J.}~\bibnamefont{Mor\'e}}, \bibinfo {author}
  {\bibfnamefont{W.}~\bibnamefont{Nazarewicz}}, \bibinfo {author}
  {\bibfnamefont{J.}~\bibnamefont{Sarich}},\ and\ \bibinfo {author}
  {\bibfnamefont{M.~V.}\ \bibnamefont{Stoitsov}},\ }%
  \bibfield{journal}{%
  \bibinfo {journal} {Phys. Rev. C}\ }%
  \textbf{\bibinfo {volume} {81}},\ \bibinfo {pages} {024316} (\bibinfo {year}
  {2010})%
  \bibAnnoteFile{NoStop}{Sch10}%
\bibitem{Pot10a}%
  \BibitemOpen
  \bibfield{author}{%
  \bibinfo {author} {\bibfnamefont{K.~J.}\ \bibnamefont{Pototzky}}, \bibinfo
  {author} {\bibfnamefont{J.}~\bibnamefont{Erler}}, \bibinfo {author}
  {\bibfnamefont{P.-G.}\ \bibnamefont{Reinhard}},\ and\ \bibinfo {author}
  {\bibfnamefont{V.~O.}\ \bibnamefont{Nesterenko}},\ }%
  \bibfield{journal}{%
  \bibinfo {journal} {Eur. Phys. J. A}\ }%
  \textbf{\bibinfo {volume} {46}},\ \bibinfo {pages} {299} (\bibinfo {year}
  {2010}),\ \url{http://dx.doi.org/10.1140/epja/i2010-11045-6}%
  \bibAnnoteFile{NoStop}{Pot10a}%
\bibitem{Pre92aB}%
  \BibitemOpen
  \bibfield{author}{%
  \bibinfo {author} {\bibfnamefont{W.~H.}\ \bibnamefont{Press}}, \bibinfo
  {author} {\bibfnamefont{S.~A.}\ \bibnamefont{Teukolsky}}, \bibinfo {author}
  {\bibfnamefont{W.~T.}\ \bibnamefont{Vetterling}},\ and\ \bibinfo {author}
  {\bibfnamefont{B.~P.}\ \bibnamefont{Flannery}},\ }%
  \emph{\bibinfo {title} {Numerical Recipes in {C}: {T}he Art of Scientific
  Computing}},\ \bibinfo {edition} {2nd}\ ed.\ (\bibinfo {publisher} {Cambridge
  University Press},\ \bibinfo {address} {New York},\ \bibinfo {year} {1992})%
  \bibAnnoteFile{NoStop}{Pre92aB}%
\bibitem{Reinhard_2013}%
  \BibitemOpen
  \bibfield{author}{%
  \bibinfo {author} {\bibfnamefont{P.-G.}\ \bibnamefont{Reinhard}}\ and\
  \bibinfo {author} {\bibfnamefont{W.}~\bibnamefont{Nazarewicz}},\ }%
  \bibfield{journal}{%
  \bibinfo {journal} {Phys. Rev. C}\ }%
  \textbf{\bibinfo {volume} {87}},\ \bibinfo {pages} {014324} (\bibinfo {year}
  {2013})%
  \bibAnnoteFile{NoStop}{Reinhard_2013}%
\bibitem{Haver17}%
  \BibitemOpen
  \bibfield{author}{%
  \bibinfo {author} {\bibfnamefont{T.}~\bibnamefont{Haverinen}}\ and\ \bibinfo
  {author} {\bibfnamefont{M.}~\bibnamefont{Kortelainen}},\ }%
  \bibfield{journal}{%
  \bibinfo {journal} {J. Phys. G}\ }%
  \textbf{\bibinfo {volume} {44}},\ \bibinfo {pages} {044008} (\bibinfo {year}
  {2017}),\ \url{http://dx.doi.org/10.1088/1361-6471/aa5e07}%
  \bibAnnoteFile{NoStop}{Haver17}%
\bibitem{Lattimer2014}%
  \BibitemOpen
  \bibfield{author}{%
  \bibinfo {author} {\bibfnamefont{J.~M.}\ \bibnamefont{Lattimer}}\ and\
  \bibinfo {author} {\bibfnamefont{A.~W.}\ \bibnamefont{Steiner}},\ }%
  \bibfield{journal}{%
  \Doi{10.1140/epja/i2014-14040-y}{\bibinfo {journal} {Eur. Phys. J. A}}\ }%
  \textbf{\bibinfo {volume} {50}},\ \bibinfo {pages} {40} (\bibinfo {year}
  {2014}),\ ISSN \bibinfo {issn} {1434-601X},\
  \url{http://dx.doi.org/10.1140/epja/i2014-14040-y}%
  \bibAnnoteFile{NoStop}{Lattimer2014}%
\bibitem{Oertel17}%
  \BibitemOpen
  \bibfield{author}{%
  \bibinfo {author} {\bibfnamefont{M.}~\bibnamefont{Oertel}}, \bibinfo {author}
  {\bibfnamefont{M.}~\bibnamefont{Hempel}}, \bibinfo {author}
  {\bibfnamefont{T.}~\bibnamefont{Kl\"ahn}},\ and\ \bibinfo {author}
  {\bibfnamefont{S.}~\bibnamefont{Typel}},\ }%
  \bibfield{journal}{%
  \Doi{10.1103/RevModPhys.89.015007}{\bibinfo {journal} {Rev. Mod. Phys.}}\ }%
  \textbf{\bibinfo {volume} {89}},\ \bibinfo {pages} {015007} (\bibinfo {month}
  {Mar}\ \bibinfo {year} {2017}),\
  \url{http://link.aps.org/doi/10.1103/RevModPhys.89.015007}%
  \bibAnnoteFile{NoStop}{Oertel17}%
\bibitem{Sha93b}%
  \BibitemOpen
  \bibfield{author}{%
  \bibinfo {author} {\bibfnamefont{M.~M.}\ \bibnamefont{Sharma}}, \bibinfo
  {author} {\bibfnamefont{G.~A.}\ \bibnamefont{Lalazissis}},\ and\ \bibinfo
  {author} {\bibfnamefont{P.}~\bibnamefont{Ring}},\ }%
  \bibfield{journal}{%
  \bibinfo {journal} {Phys. Lett.}\ }%
  \textbf{\bibinfo {volume} {B317}},\ \bibinfo {pages} {9} (\bibinfo {year}
  {1993})%
  \bibAnnoteFile{NoStop}{Sha93b}%
\bibitem{Rei95a}%
  \BibitemOpen
  \bibfield{author}{%
  \bibinfo {author} {\bibfnamefont{P.-G.}\ \bibnamefont{Reinhard}}\ and\
  \bibinfo {author} {\bibfnamefont{H.}~\bibnamefont{Flocard}},\ }%
  \bibfield{journal}{%
  \bibinfo {journal} {Nucl. Phys.}\ }%
  \textbf{\bibinfo {volume} {A584}},\ \bibinfo {pages} {467} (\bibinfo {year}
  {1995})%
  \bibAnnoteFile{NoStop}{Rei95a}%
\bibitem{Helm56}%
  \BibitemOpen
  \bibfield{author}{%
  \bibinfo {author} {\bibfnamefont{R.~H.}\ \bibnamefont{Helm}},\ }%
  \bibfield{journal}{%
  \Doi{10.1103/PhysRev.104.1466}{\bibinfo {journal} {Phys. Rev.}}\ }%
  \textbf{\bibinfo {volume} {104}},\ \bibinfo {pages} {1466} (\bibinfo {month}
  {Dec}\ \bibinfo {year} {1956}),\
  \url{https://link.aps.org/doi/10.1103/PhysRev.104.1466}%
  \bibAnnoteFile{NoStop}{Helm56}%
\bibitem{Mizutori00}%
  \BibitemOpen
  \bibfield{author}{%
  \bibinfo {author} {\bibfnamefont{S.}~\bibnamefont{Mizutori}}, \bibinfo
  {author} {\bibfnamefont{J.}~\bibnamefont{Dobaczewski}}, \bibinfo {author}
  {\bibfnamefont{G.~A.}\ \bibnamefont{Lalazissis}}, \bibinfo {author}
  {\bibfnamefont{W.}~\bibnamefont{Nazarewicz}},\ and\ \bibinfo {author}
  {\bibfnamefont{P.-G.}\ \bibnamefont{Reinhard}},\ }%
  \bibfield{journal}{%
  \Doi{10.1103/PhysRevC.61.044326}{\bibinfo {journal} {Phys. Rev. C}}\ }%
  \textbf{\bibinfo {volume} {61}},\ \bibinfo {pages} {044326} (\bibinfo {month}
  {Mar}\ \bibinfo {year} {2000}),\
  \url{https://link.aps.org/doi/10.1103/PhysRevC.61.044326}%
  \bibAnnoteFile{NoStop}{Mizutori00}%
\bibitem{Yam12}%
  \BibitemOpen
  \bibfield{author}{%
  \bibinfo {author} {\bibfnamefont{M.}~\bibnamefont{Yamagami}}, \bibinfo
  {author} {\bibfnamefont{J.}~\bibnamefont{Margueron}}, \bibinfo {author}
  {\bibfnamefont{H.}~\bibnamefont{Sagawa}},\ and\ \bibinfo {author}
  {\bibfnamefont{K.}~\bibnamefont{Hagino}},\ }%
  \bibfield{journal}{%
  \Doi{10.1103/PhysRevC.86.034333}{\bibinfo {journal} {Phys. Rev. C}}\ }%
  \textbf{\bibinfo {volume} {86}},\ \bibinfo {pages} {034333} (\bibinfo {month}
  {Sep}\ \bibinfo {year} {2012}),\
  \url{https://link.aps.org/doi/10.1103/PhysRevC.86.034333}%
  \bibAnnoteFile{NoStop}{Yam12}%
\bibitem{Kor12a}%
  \BibitemOpen
  \bibfield{author}{%
  \bibinfo {author} {\bibfnamefont{M.}~\bibnamefont{Kortelainen}}, \bibinfo
  {author} {\bibfnamefont{J.}~\bibnamefont{McDonnell}}, \bibinfo {author}
  {\bibfnamefont{W.}~\bibnamefont{Nazarewicz}}, \bibinfo {author}
  {\bibfnamefont{P.-G.}\ \bibnamefont{Reinhard}}, \bibinfo {author}
  {\bibfnamefont{J.}~\bibnamefont{Sarich}}, \bibinfo {author}
  {\bibfnamefont{N.}~\bibnamefont{Schunck}}, \bibinfo {author}
  {\bibfnamefont{M.~V.}\ \bibnamefont{Stoitsov}},\ and\ \bibinfo {author}
  {\bibfnamefont{S.~M.}\ \bibnamefont{Wild}},\ }%
  \bibfield{journal}{%
  \bibinfo {journal} {Phys. Rev. C}\ }%
  \textbf{\bibinfo {volume} {85}},\ \bibinfo {pages} {024304} (\bibinfo {year}
  {2012})%
  \bibAnnoteFile{NoStop}{Kor12a}%
\end{thebibliography}%

\end{document}